\documentclass[12pt]{iopart}
\usepackage{graphicx}
\usepackage{amsmath}
\usepackage{amssymb}
\usepackage{multirow, makecell, comment} 
\newcommand{\fla}[1]{\begin{flalign}#1\end{flalign}}
\usepackage{braket}
\usepackage{lineno}
\usepackage{lipsum}
\usepackage{color}
\usepackage{xcolor}
\usepackage{soul}
\usepackage{array} 
\usepackage{cite}
\usepackage{xcolor} 
\usepackage[makeroom]{cancel}

\usepackage{xcolor} %
\usepackage[english]{babel} 
\usepackage{iopams}  
\begin{document}

\title[Echo protocols of an optical quantum memory]{Echo protocols of an optical quantum memory}

\author{S.A. Moiseev$^{1*}$, K.I. Gerasimov$^1$, M.M. Minnegaliev$^1$, E.S. Moiseev$^1$, A.D. Deev$^{2,3}$, Yu.Yu. Balega$^4$}
\address{$^1$ Kazan Quantum Center, Kazan National Research Technical University, 10 K. Marx, Kazan, 420111, Russia}
\address{$^2$ 
 Moscow Institute of Physics and Technology
117303, Moscow, Kerchenskaya Street, 1A, building 1.}
\address{$^3$ 
"R\&D Center" JSC,
 Bolshoy Balkansky Lane 20, 129090 Moscow, Russian Federation}
\address{$^4$ Russian Academy of Sciences, China Branch of BRICS Institute of Future Networks (Shenzhen, China).}

\ead{s.a.moiseev@kazanqc.org $^{*}$}

\vspace{10pt}
\begin{indented}
\item[]April 2024
\end{indented}

\begin{abstract}
Based on new obtained analytical results, the main properties of photon echo quantum memory protocols are analysed and discussed together with recently achieved experimental results. 
The main attention is paid to studying the influence of spectral dispersion and nonlinear interaction of light pulses with resonant atoms.
The distinctive features of the effect of spectral dispersion on the quantum storage of broadband signal pulses in the studied echo protocols are identified and discussed.
Using photon echo area theorem, closed  analytical solutions for echo protocols of quantum memory are obtained, describing the storage of weak and intense signal pulses, allowing us to find the conditions for the implementation of high efficiency in the echo protocols under strong nonlinear interaction of signal and control pulses with atoms. 
The key existing practical problems and the ways to solve them in realistic experimental conditions are outlined.
We also briefly discuss the potential of using the considered photon echo quantum memory protocols in a quantum repeater.
\end{abstract}

%\includegraphics[scale=0.5]{GA-final.pdf}

%42.25.Bs Wave propagation, transmission and absorption
%42.79.Gn Optical waveguides and couplers
%41.20.Jb Electromagnetic wave propagation; radiowave propagation%79.60.Jv Interfaces; heterostructures; nanostructures
%78.67.Pt Multilayers; superlattices; photonic structures; metamaterials

\vspace{2pc}
\noindent{\it Keywords}: Optical quantum memory, photon echo,  crystals with rare earth ions, quantum repeater. 
%\maketitle
\ioptwocol

\section{Introduction}

Optical quantum memory (QM) is a device that is designed to store the quantum state of light, e.g., photonic qubits, for a given time. 
Optical QM is considered as a crucial element in many quantum technologies, for example, in a quantum repeater\cite{Sangouard2011}, a universal optical quantum computer \cite{Kok2007}, quantum light sources \cite{Stevenson2014,Manukhova2017, Chen2023} as well as in experiments on fundamental tests of quantum theory \cite{Clauser1969,Brunner2014,Mol2023}. 
The optical QM must provide high efficiency (low energy losses), high fidelity of the recalled quantum states, a sufficiently long lifetime and high information capacity. 
Practically important physical parameters of the QMs are the operating wavelength of light fields and the possibility of integration into existing devices. 
Performing additional operations with quantum states of light \cite{Hosseini2012} and quantum addressing \cite{Moiseev2016, Chen2021} is becoming increasingly important. 

In the last two decades, significant progress has been achieved in the development of optical QM based on the photon echo effect \cite{Lvovsky2009, Tittel2009,Bussieres2013,Heshami2016a,Chaneliere2018,Hua2018,Guo2023,Zhou2023,Lei2023}.
The interest in echo-protocols QM is caused by their ability to store a large number of photonic qubits in a single QM cell. 
Such a multi-mode QM has great versatility, allowing quantum storage of both single-photon and intense light fields in arbitrary quantum states, while preserving their quantum properties, which is of great importance for quantum communications and quantum computing.

Generalizing the ideas of spin echo \cite{Hahn1950}  and photon echo \cite{Kopvillem1963, Kurnit1964}, echo-based QM protocols ensure the reversibility of quantum dynamics not only of an ensemble of spins or atoms, but also of signal fields resonantly interacting with them.
Conceptually, the full reversal of the quantum dynamics in an atom-field system  makes the echo based QM  protocols a variant of the Loschmidt echo 
\cite{GORIN2006,Wisniacki2012}  with a key difference in presence of an intermediate state in which an input light pulse is completely absorbed by atoms.
The possibility of implementing such a variant of photon echo was proposed and justified in \cite{Moiseev2001, Moiseev2004,Moiseev2003,Moiseev2004a}  for an optical depth atomic ensemble with \textit{controlled reversion of inhomogeneous broadening} (CRIB) of the resonant atomic detunings called later by the CRIB protocol in \cite{Kraus2006,Alexander2006,Lauritzen2010} and then used in the development of modified echo protocols adaptable for easier practical implementation \cite{Tittel2009,Chaneliere2018,Heshami2016a}.

In this article, based on obtaining a number of new analytical results, we analyze the basic physical properties and conditions for the effective implementation of photon echo QM  protocols of weak and intense light fields, discover  their  specific unknown  features, advantages and difficulties in practical implementation. 
Using the concepts of the developed theory, the most important experimental results obtained in the studied QM echo protocols and the existing problems in achieving  basic  parameters are also discussed.

Following the original works of the CRIB protocol \cite{Moiseev2001, Moiseev2004,Moiseev2003,Moiseev2004a}  and its further generalizations \cite{Kraus2006, Sangouard2007,Gorshkov2007,Moiseev2011,Moiseev2013},
we first consider the basic scenario and  physical properties of the QM echo protocol.
Then we study the physical features that arise in its modified versions caused by the use of various inhomogeneous broadening of the atomic transition and methods of controlling the excited optical quantum coherence.
Here we show new manifestations of spectral dispersion and the resulting problems in the implementation of  broadband QM echo protocols and their relationship to the presence or absence of temporal reversibility, and discuss ways to solve the problems that have arisen. 
The analysis is summarized by discussing the properties of the echo protocol while storage of intense signal pulses.
Nonlinear patterns in the storage of signal pulses are studied on the basis of obtaining closed  analytical solutions for the pulse area of the stored signals, which allows us to find more general conditions for achieving the maximum efficiency of the echo-protocols.
Analyzing the properties of the studied QM echo protocols, we focus on the best experimental results achieved and ways of its further improvement.
At the end of the article, we briefly discuss the potential of using the considered photon echo quantum memory protocols in a quantum repeater. 

\section*{CRIB/GEM protocol}

The CRIB protocol proposed and justified in \cite{Moiseev2001, Moiseev2004,Moiseev2003,Moiseev2004a}  is based on the use of an ensemble of $N$ three-level atoms having an inhomogeneous broadening $G(\Delta/\Delta_{in})$ with a linewidth $\Delta_{in}$ at optical transition $\ket{1}\rightarrow \ket{3}$. 
All atoms are prepared in the state $\ket{1}$. 

\begin{figure}
%\begin{center}\vspace{1cm}
%\includegraphics[width=1.0\linewidth]{Fig1-CRIB.png}
\includegraphics[width=1\linewidth]{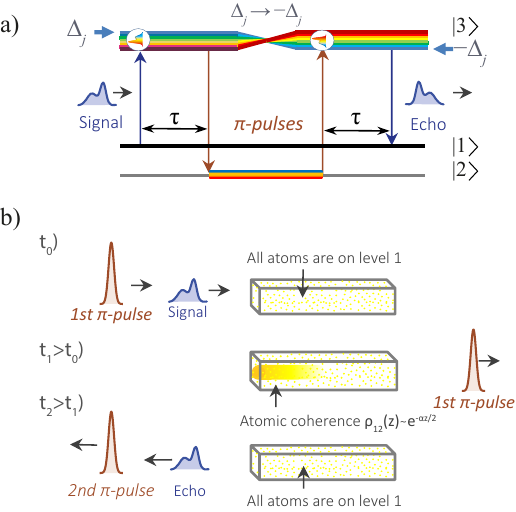}
\caption{(a) Atomic energy levels, a time-reversible scheme and (b) a spatial (backward) geometry of  CRIB protocol.}
\label{CRIB}
\end{figure}

Fig.\ref{CRIB} shows a diagram of atomic levels, the time sequence of the signal and two controlling laser $\pi$ pulses followed by photon echo emission in backward geometry.
The implementation of optical QM with intermediate storage of information about the signal pulse on the long-lived spin coherence of the atomic ground state is called the spin-wave protocol.
At stage of echo emission, the initial frequency detuning $\Delta$ of  atoms at the atomic transition $\ket{1}\leftrightarrow\ket{3}$ is reversed $-\Delta$ to restore the macroscopic atomic coherence excited by the  signal pulse.
In the first works, the CRIB-protocol  was proposed to be implemented using the Doppler effect in a gas \cite{Moiseev2001} and controlling hyperfine interactions of  rare earth ions (REI) by changing the orientation of the neighboring nuclear spins \cite{Moiseev2003}.
Experimentally, it is convenient to implement CRIB-protocol  in crystal with REIs  by creating a narrow line within inhomogeneous broadening resonant transition by using a laser hole burning technique \cite{Nilsson2004}.
The Stark (Zeeman) effect can then be applied in an external electric (magnetic) field, the gradient of which is oriented perpendicular to the propagation of the signal light field and to change its polarity on demand\cite{Alexander2006,Hosseini2011}.

The signal light pulse is launched into the QM cell with a carrier frequency that is resonant with the center of the atomic transition line $\ket{1}\rightarrow\ket{3}$. 
The signal duration $\delta t_s$ is assumed to be shorter than the phase relaxation time of the optical transition ($\delta t_s\ll\gamma=1/T_2$), and its spectral width is less than the inhomogeneous broadening of the optical transition $\delta\omega_s\sim\delta t_s^{-1}<\Delta_{in}$.
Weak signal pulses does not significantly change the population difference of atomic states at the resonant transition $\ket{1}\rightarrow\ket{3}$.
Taking this into account, the quantum Maxwell-Bloch (MB)  equations are linearized. 
The equations for slowly varying atomic coherence $\sigma^j(\tau)$ and field annihilation operator   $a_s (Z,\tau)$ 
($[\hat{a}_s(Z,\tau),\hat{a}^{\dagger}_s(Z',\tau)]=\delta(z-z')$) 
in the co-moving coordinate system $Z=z$, $\tau=t-z/v_g$ ($v_g$ is the group velocity) have the form:
\fla{
&v_g \frac{\partial}{\partial Z} \hat{a}_s(Z,\tau) =  
\frac{i}{2} \sum^{N}_{j=1} \Omega (\mathbf{r}^{j}_{\perp} ) \hat{\sigma}^{j}_{-}\delta(Z-Z_j),
\nonumber
\\
%%%%%%%%%%%%%%%%%%%%%%%%%%%%%%%%%
& \frac{\partial \hat{\sigma}^{j}_{-} }{\partial \tau} = - i\Delta_{j}  \hat{\sigma}^{j}_{-} 
+\frac{i}{2} \Omega^{*}(\mathbf{r}^{j}_{\perp})\hat{a}_p(Z_j,\tau),
\label{eq-1}
}
where $\Delta_j$ is a frequency offset of j-th atom.
In Eq.\eqref{eq-1} the effect of relaxation and Langevin forces during interaction with the signal pulse are neglected. 
By introducing the interaction constant $\Omega(\mathbf{r}^{j}_{\perp})$, we also included echo protocols in a single-mode waveguide \cite{Moiseev2023}
($\Omega(\mathbf{r}_{\perp}^j )=\Omega_{0} f(\mathbf{r}_{\perp}^j)$ is a  coupling constant between $j$-th atom and waveguide mode with 
$\Omega_{0}=E_0 \langle \mathbf{d}\cdot \mathbf{e} \rangle/\hbar$
being a single photon Rabi-frequency of  $j$-th atom located at $\mathbf{r}^{j}$. 
The membrane function $f(\mathbf{r}_{\perp})$ determines the dependence of the interaction constant of an atom on its position $\mathbf{r_{\perp}}$ in the transverse plane of the waveguide.

The solution of Eq.\eqref{eq-1} describes the resonant absorption by atoms of a signal field, the Fourier components of which are attenuated by Behr's law 
$ \hat{a}_s(Z,\omega)=\hat{a}_s(Z,0)e^{-\alpha(\omega) Z/2}$ with an absorption coefficient 

\fla{
\alpha(\omega)=\beta\chi(\omega)=
\beta\int d\delta \frac{G(\delta/\Delta_{in})}{[1/T_2+i(\delta-\omega)}, 
\label{absoprion_coef}
}

\noindent
where 
$\beta=\frac{N\langle \Omega(\mathbf{r}_{\perp})|^2\rangle}{2Lv_g}$, 
$N$ is a number of atoms, 
$\langle |\Omega(\mathbf{r}_{\perp})|^2\rangle=\frac{1}{S}\iint |\Omega(\mathbf{r}_{\perp})|^2 dxdy$, $S$ and $L$ are the cross-section area and the length of the waveguide, $T_2$ is a time of phase relaxation.
For example, for the Lorentzian shape of the line $G(\delta/\Delta_{in})\rightarrow G_L (\delta/\Delta_{in})=\frac{\delta_{in}}{\pi(\Delta_{in}^2+ \delta^2)}$ 
and $\chi(\omega)=\chi_{L}(\omega)=\frac{1}{(\Delta_{in}-i\omega)}$ where $\alpha(\omega)$ is characterized by a spectrally dependent absorption coefficient and dispersion.

After an absorption of the signal ($t>\delta t_s$), the excited optical coherence of atoms passes into free evolution 

\fla{
\hat{\sigma}^{j}_{-}(\tau) =i\sqrt{\pi/2}  \Omega^{*}(\mathbf{r}^{j}_{\perp}) \hat{a}_s(0,\Delta_j)e^{-\alpha(\Delta_j)Z_j-i\Delta_j\tau}.
\label{free-evolution}
}
This behavior demonstrates the characteristic properties of QM echo protocols: the spectral components of light pulses are recorded in the corresponding spectral components of atomic coherence and the condition $Re\{\alpha(\Delta)\}L\gg1$ determines the maximum achievable spectral width of the QM.
The spatial distribution of coherence $\hat{\sigma}^{j}_{-}(\tau>\delta t_s)$ does not depend on the temporal shape of the signal pulse.
The signal field can contain a large number of pulses, which is limited only by the ratio $\Delta_{in}/\gamma \gg 1$. 
However, short signal pulses ($\delta\omega_s\sim\Delta_{in}$) 
acquire undesirable phase modulation in the depth of the medium due to the growing influence of spectral dispersion $\alpha(\Delta)$.
%begin to experience undesirable phase modulation due to the growing influence of the spectral dispersion $\alpha(\Delta)$ in an optical depth medium.
Below, we discuss the negative effect of the spectral dispersion on the properties of echo protocols and the possibilities of its suppression.

To increase the QM storage time, it is common to apply a short control laser $\pi$ pulse that propagates parallel to the signal pulse $\mathbf{k}_1\uparrow\uparrow \mathbf{k}_s$ 
%(if $t_1>\delta t_s$) 
($\delta t_1\ll\delta t_s$) and being in resonance with the frequency of the atomic transition $\ket{2}\leftrightarrow \ket{3}$ \cite{Moiseev2001}. 
The $\pi$ pulse, without experiencing absorption, converts the optical coherence $\hat{\sigma}^{j}_{-}(t_1)$ at time $t_1$ to the long-lived spin coherence  $\hat{\sigma}^{j}_{12}(t_1)$ between transition $\ket{1}\rightarrow \ket{2}$  
which completes the mapping of the quantum state of the signal pulse into the  QM cell.
During the storage time T, spin coherence is negatively affected by static and dynamic local fluctuations of the magnetic field, dipole-dipole interactions with neighboring electron and nuclear spins, which lead to irreversible decay  in spin coherence. 
Highly efficient suppression of these effects can be achieved by using additional rephasing  radio-frequency pulses \cite{Merkel2021, Moiseev2015,Moiseev2015PRA,Waeber2019, Minnegaliev2019}, as well as placing the crystal in an external magnetic field \cite{Heinze2013, Zhong2015, Rancic2017}, in which the frequencies of spin transitions are not sensitive to weak local fluctuations of the magnetic field. 
These methods open up ways to create QM with a lifetime reaching minutes and hours \cite{Hain2022, Ma2021}.

In the CRIB protocol, before applying a short second laser $\pi$ pulse ($\delta t_2\ll\delta t_s$), the frequency detuning of each atom changes to the opposite ($\Delta_j\rightarrow-\Delta_j$). 
The second $\pi$ pulse propagates in the opposite direction to the signal pulse 
($\mathbf{k}_2\uparrow\downarrow \mathbf{k}_s$ ) 
and, being resonant to the atomic the transition $\ket{2}\rightarrow \ket{3}$, transfers the spin coherence $\hat{\sigma}^{j}_{12}(t_2)$  at time $t_2=T+t_1$ to the optical coherence $\hat{\sigma}^{j}_{-}(t_2)$ of the atomic transition $\ket{1}\rightarrow \ket{3}$.
At the time $t_e=T+2t_1$, the coherence of atoms is phasing when 

\fla{
\hat{\sigma}^{j}_{-}(t) =&-i\sqrt{\pi/2}  \Omega^{*}(\mathbf{r}^{j}_{\perp}) \hat{a}_s(0,\Delta_j)\cdot
\nonumber
\\
&e^{-\alpha(\Delta_j)Z+i\Delta_j(t-t_e)},
\nonumber
}
which leads to the generation of a photon echo  in the opposite direction to the signal pulse $\mathbf{k}_e\uparrow\downarrow \mathbf{k}_s$ (see Fig.\ref{CRIB}).

Here we should note that in a waveguide, the exact realization of laser $\pi$ pulses becomes impossible for all the atoms located at different points of the waveguide cross-section. 
In this case, as shown in a recent work \cite{Moiseev2023}, it is still possible to obtain analytical solutions for strong nonlinear interaction of the control laser pulses with atoms when analysing  the pulse areas of the studied light pulses.
However, the QM protocol efficiency  is becoming less and we will conduct further studies  under the assumption of plane waves for the control laser pulses having the same intensity in the cross section of the signal and echo fields.
The  echo signal emission is described by a system of linear equations for slow variables of atomic coherence and amplitude of the echo signal in a moving coordinate system 
$Z=z$, $\tilde{\tau}=t+z/v_g$
The equations differ from Eq.\eqref{eq-1} only in replacing
$\frac{\partial}{\partial Z}\rightarrow - \frac{\partial}{\partial Z}$,$\Delta_j\rightarrow-\Delta_j$, and non-zero initial conditions for the atomic coherence. 
The solution is written using the Fourier components of the echo signal 
$\hat{a}_e(Z=0,t)=\frac{1}{\sqrt{2\pi}}
\int d\omega \tilde{a}_e(0,\omega )e^{i\omega (t-t_e)}$
on the output medium ($Z=0$):

\fla{
&\tilde{a}_e(0,\omega )=
\Gamma(t_e )\tilde{a}_s(0,-\omega )\cdot
\nonumber
\\
& 
\frac{2\pi\beta G(-\omega /\Delta_{in})}{[\alpha(\omega )+\alpha(-\omega )]}
\{1-e^{-[\alpha(\omega )+\alpha(-\omega )]L}\},
\label{eq-2}
}
where $\Gamma(t_e )=\langle e^{-i\delta\phi_j(t_e)}\rangle$ takes into account the phase relaxation of atomic coherence at the time of the echo emission ($\delta\phi_j(t_e)$ is a random phase acquired by a atom during the time the signal is stored in the QM cell).
Taking into account the symmetric inhomogeneous broadening, due to which 

\fla{
\frac{1}{2}[\alpha(\omega )+\alpha(-\omega )]=\pi\beta G(\omega /\Delta_{in})\equiv \alpha_R(\omega ),
\label{abs_coeff}
}
where for the Lorentzian inhomogeneous broadening $\alpha_R(\omega )\equiv\alpha_{RL}(\omega )=\beta\frac{\Delta_{in}}{\Delta_{in}^2+\omega ^2}$ is an absorption coefficient.

At the output of the medium, we get \cite{Moiseev2004,Sangouard2007b}:

\fla{
\tilde{a}_e(0,\omega )=\Gamma(t_e )\tilde{a}_s(0,-\omega )\{1-e^{-\alpha_R(\omega )L}\}.
\label{eq-3}
}

According to Eq.\eqref{eq-3}, 
the effect of spectral dispersion is fully compensated in the backward  CRIB protocol  \cite{Moiseev2004}.
In the presence of sufficient large optical depth ($\alpha_R(\omega )L\gg1$) and weak atomic phase relaxation ($\Gamma(t_e )\cong1$) the photon echo ($\tilde{a}_e(0,\omega )=\tilde{a}_s(0,-\omega )$) completely restores the signal pulse in the form of its time-reversed copy $\tilde{a}_e(0,t)=\tilde{a}_s(t_e-t)$ \cite{Moiseev2001,Moiseev2004}.
The maximum spectral width $\omega_s$  of the signal pulse  is determined only by the optical depth of the medium $\alpha_R(\delta\omega_s)L$ in Eq.\eqref{eq-3}.

It is worth noting that the absence of the influence of spectral dispersion on the properties of the CRIB protocol in the backward geometry is due to the accuracy of the reversibility of  MB equations describing the stages of signal absorption and photon echo emission when performing transformations:
$\Delta_j\rightarrow-\Delta_j$, $\tau\rightarrow-\tilde{\tau}$,
$a_s(Z,\tau)\rightarrow -a_e(Z,\tilde{\tau})$, which is valid even in the case of intense signal pulses \cite{Kraus2006,Moiseev2011}.
Using this property, it is possible to find the wave function of echo signals even in the  complex (entangled) multiphoton states of the input signal pulses \cite{Moiseev2007b}.
The use of a superposition of quantum states of the initial signal pulse and a photon echo with spectrally inverted wave function parameters provides an interesting resource for the preparation of new entangled states of multiphoton fields.
It was also shown \cite{Moiseev2013}  that using the symmetry of MB equations of the both stages in CRIB protocol on off-resonant Raman transitions makes it possible to manipulate the temporal duration and carrier frequency of echo signal.
The possibilities of the wavelength conversion  in the photon echo QM technique are also considered in the work \cite{Ham2018}.
The concept of backward geometry of echo retrieval proposed in the CRIB protocol was also applied in the effective implementation of QM based on electromagnetically induced transparency \cite{Gorshkov2007PRL}, Autler-Townes splitting \cite{Autler1955,Saglamyurek2018} and QM on off-resonant Raman scattering \cite{Vernaz-Gris2018}.
From an engineering point of view, the implementation of such geometry is not difficult. 
Similar propagation of signal and control laser fields has been experimentally implemented in such works as \cite{Dajczgewand2014,Minnegaliev2022}.

\begin{figure}
%\begin{center}\vspace{1cm}
%\includegraphics[width=1\linewidth]{Fig2_crib_ab_F.png}
%\includegraphics[width=1\linewidth]{Fig2_crib_ab_F.eps} 
\includegraphics[width=1\linewidth]{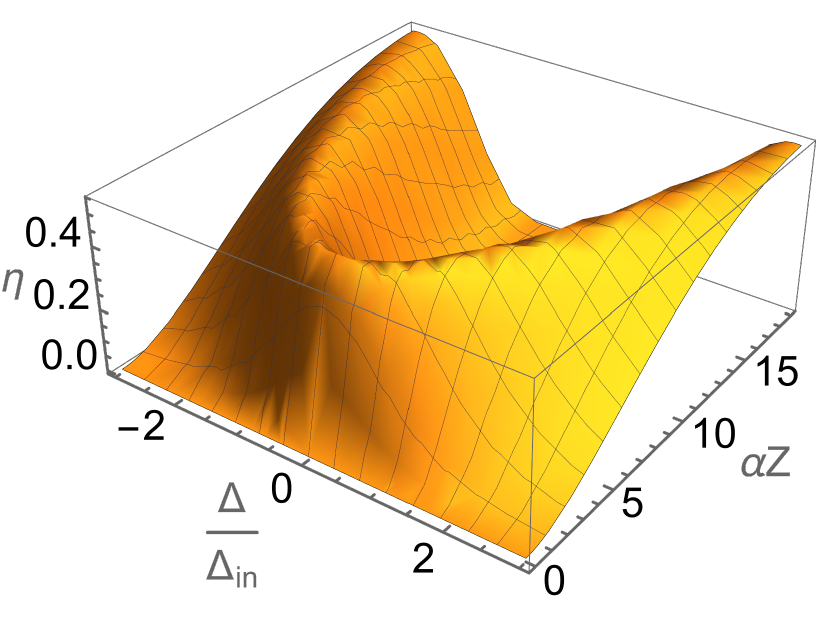}
\caption{Spectral efficiency $\eta$ for storage of narrowband signal pulse described by Eq. \eqref{sp_eff_CRIB_ansorb_F} in forward geometry of CRIB protocol as function of optical depth $\alpha L$ and spectral offset  $\Delta/\Delta_{in}$.}
\label{spectral-eff-CRIB-abs}
\end{figure}

If the $\pi$-pulses of equal wave vectors 
$\mathbf{k}_1\uparrow\uparrow \mathbf{k}_2$ are applied,
then the echo signal will be emitted in the same direction
$\mathbf{k}_e\uparrow\uparrow \mathbf{k}_s$. 
In this forward geometry, the solution of the MB equations for the echo signal in the medium has the form
$\hat{a}_e(Z,t)=\frac{1}{\sqrt{2\pi}}\int d\omega  \tilde{a}_e(Z,\omega )e^{i\omega (t-t_e-Z/v_g)}$, where:

\fla{
&\tilde{a}_e(Z,\omega )=
\Gamma(t_e )\tilde{a}_s(0,-\omega )\cdot
\nonumber
\\
& 
\frac{2\pi G(-\omega /\Delta_{in})}{[\chi(\omega )-\chi(-\omega )]}
\{e^{-\alpha(-\omega  )Z/2}-e^{-\alpha(\Delta)Z/2}\}.
\label{eq-4}
}

\noindent
Considering  Lorentzian   inhomogeneous broadening in \eqref{eq-4}, we get: 

\fla{
\tilde{a}_e(Z,\omega )=&
\Gamma(t_e )\tilde{a}_s(0,-\omega a)\cdot
\nonumber
\\
& 
\frac{\sin[\frac{\omega}{2\Delta_{in}}\alpha_{RL}(\omega)Z] }{\omega/(2\Delta_{in}) }
e^{-\alpha_{RL}(\omega)Z/2}.
\label{eq-5}
}
According to \eqref{eq-5}, the spectral dispersion contributes to a decrease in spectral efficiency 
$\eta(\omega,Z)=\frac{\langle I_e(Z,\omega)\rangle}{\langle I_e(0,\omega)\rangle}$ (where $\langle I_{e,s}(z,\omega)\rangle=\langle\hat{a}_{e,s}^{\dagger}(z,\omega)
\hat{a}_{e,s}(z,\omega)
\rangle$) as follows:

\fla{
\eta(\omega,Z)\sim
e^{-\alpha_{RL}(\omega)Z}
\frac{\sin^2[\frac{\omega}{2\Delta_{in}}\alpha_{RL}(\omega)Z] }{[\omega/(2\Delta_{in})]^2}.
\label{eq-6}
}

Note that the exponential factor in the expression \eqref{eq-6} reflects the effect of the absorption coefficient, and the second factor describes the influence of spectral dispersion.
For narrowband signal pulse $\frac{\delta\omega_s}{\Delta_{in}}<0.1$, the maximum  $\eta_m(\omega)$ reaches at  $\alpha_R(\omega)L\approx 2$  and Eq.\eqref{abs_coeff} reduces to the well-known solution for $\frac{\omega}{2\Delta_{in}}\alpha_R(\omega)Z\ll\pi/2$:

\fla{
\tilde{a}_e(0,\omega)\cong
\Gamma(t_e )\tilde{a}_s(0,-\omega)\alpha_R(\omega)Z 
e^{-\alpha_R(\omega)Z/2}. 
\label{sp_eff_CRIB_ansorb_F}
}

\begin{figure}
%\begin{center}\vspace{1cm}
\includegraphics[width=1\linewidth]{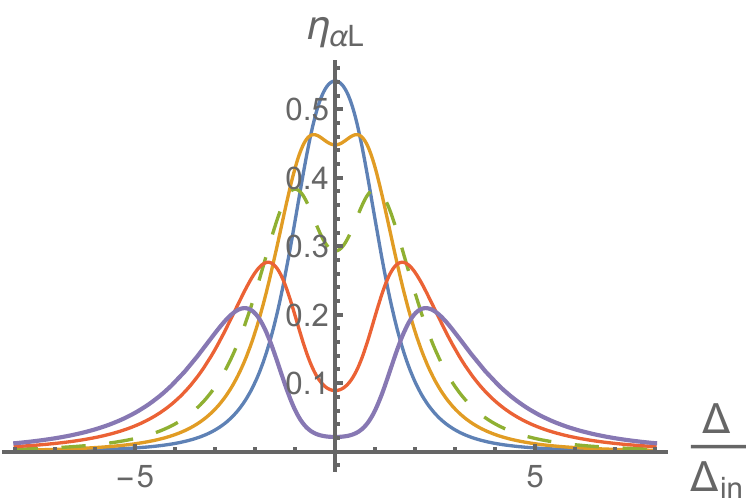}
\includegraphics[width=1\linewidth]{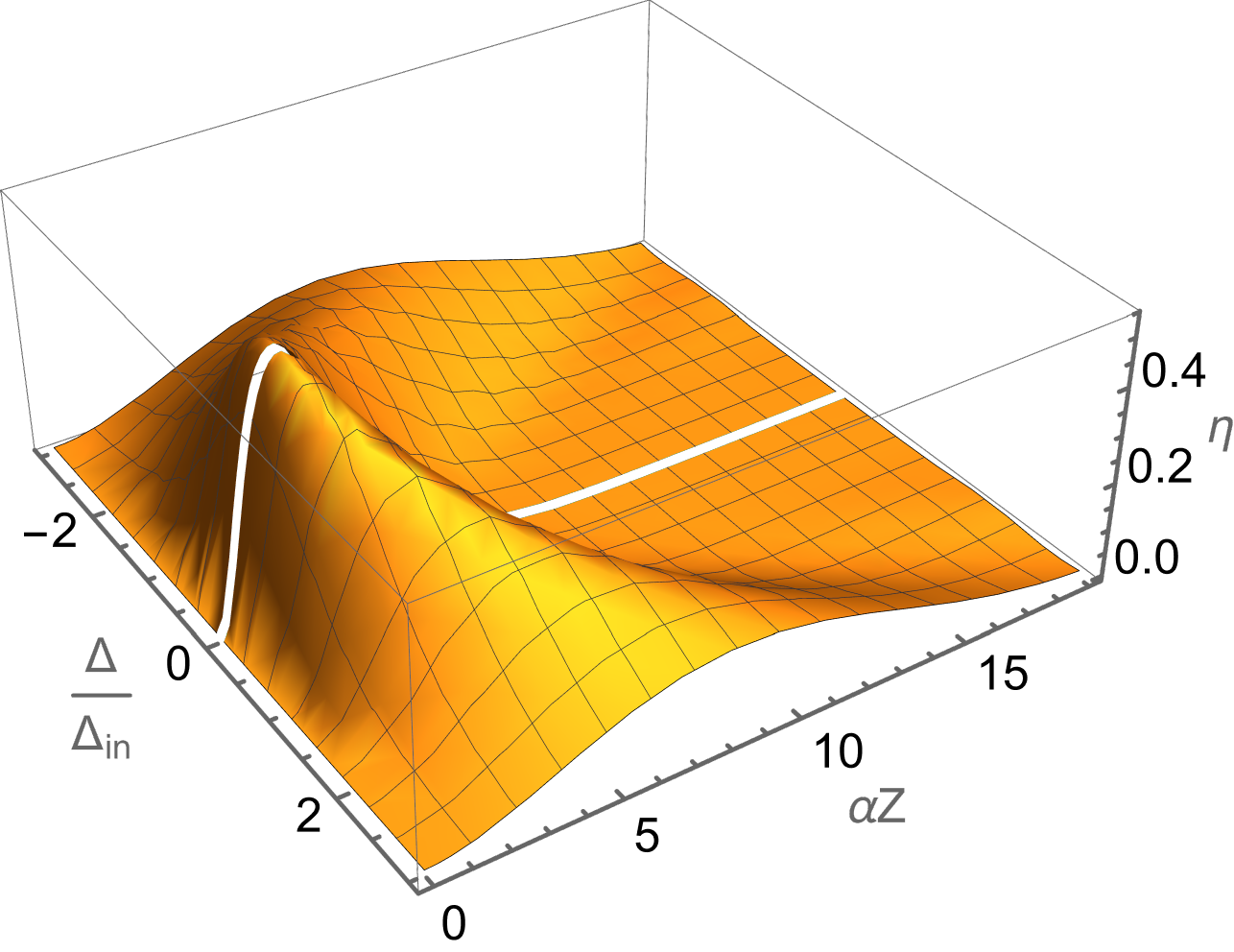}
\caption{ Spectral efficiency $\eta_{\alpha_R(0)L}$ of the forward CRIB-protocol \eqref{eq-6}; 
top: at different optical depth in the center of the absorption line ($\eta_2$(blue), $\eta_3$(dashed), $\eta_4$ (dashing[large]), $\eta_6$ (dotdashed), $\eta_8$ (thick)) (where 
$\alpha_R(0) L=n$ for $\eta_n$);
bottom: spectral efficiency as a function of the optical depth $\alpha L$ and spectral offset  $\Delta/\Delta_{in}$.}
\label{CRIB-SpectrEff}
\end{figure}

\noindent
where the spectral efficiency no longer depends on the dispersion of the medium, but is determined only by the spectral properties of the absorption coefficient
$\eta(\omega)\sim \alpha_R^2(\omega)Z^2 
e^{\alpha_R(\omega)Z}$
shown in the Fig.\ref{spectral-eff-CRIB-abs}.
%The spectral efficiency  of the forward CRIB protocol is  determined only by the spectral behavior of the absorption coefficient (see Fig.\ref{spectral-eff-CRIB-abs}).
The maximum of spectral efficiency  $\eta(\omega)$  reaches the same value of 54\%, but at the optimal distance $L(\omega)=2/\alpha_R(\omega)$ that increases with increasing  of the frequency detuning $\frac{|\omega|}{\Delta_{in}}$.
The spread of optimal distances reduces the overall efficiency of the protocol while storage signal fields with a larger spectral width.
However, the spectral efficiency of storing broadband signals is even more influenced by spectral dispersion.
The effect of spectral dispersion on spectral efficiency depending on the optical depth of the medium is shown  
%Spectral dispersion strongly affects the spectral properties of efficiency depending on its optical depth as it is depicted 
in top graphic images of Fig.\ref{CRIB-SpectrEff} ($\eta_n: \alpha_R(0) L=n)$).
The efficiency $\eta(\omega,Z)$  reaches its maximum  $\eta_m(\omega,L)$ at optical depth 
$\alpha_{RL}(\omega)L=\frac{\Delta_{in}}{\omega}
\arctan{(\frac{2\omega}{\Delta_{in}})}$ 
where
$\eta_m(\omega,L)=\frac{4}{1+4(\frac{\omega}{\Delta_{in}})^2}\cdot$
$\exp\{-\alpha_{RL}(\omega)L\}$. 

\begin{figure}
%\begin{center}
%\vspace{1cm}
\centering
\includegraphics[width=1.0\linewidth]{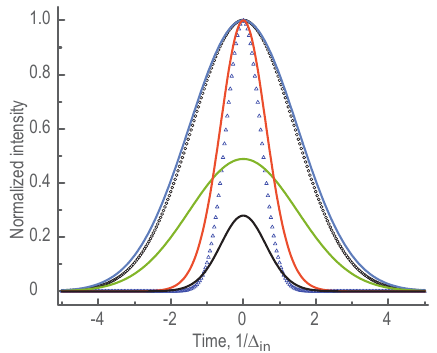}
\caption{Gaussian input pulses with energy spectral widths (HWe$^{-1}$M) of 0.7$\cdot\Delta_{in}$ (black circles) and 1.5$\cdot\Delta_{in}$  (blue triangles) and the corresponding echo signals normalized to the maximum (blue and red curves), showing an increase in echo pulse duration with increasing input pulse spectral width. The echo signals for these two input pulses (green and black curves) are normalized to the input pulse intensity and show the decreasing efficiency with increasing input pulse spectral width.}
\label{CRIB-Spectral width}
\end{figure}

In Fig.\ref{CRIB-SpectrEff}  it is evident that the influence of spectral dispersion also leads to a decrease in the maximum spectral efficiency with increasing optical depth $\alpha_R(0) L$.
Accordingly, due to the influence of spectral dispersion, the storage of short light pulses will be also accompanied by their stronger attenuation and an additional narrowing of their spectrum (an increase in their durations) (see Fig.\ref{CRIB-Spectral width}).
Thus, unlike the backward geometry, spectral dispersion strongly affects the efficiency and fidelity of the implementation of the forward CRIB protocol. 
It should be noted that the negative manifestation of spectral dispersion in the quantum storage of broadband signal pulses is characteristic of all QM echo-protocols that do not have full temporal reversibility of the interaction of light pulses with resonant media.  
This problem becomes especially important when using optically depth media.

\textit{Nonlinear effects.} To study the efficiency of  CRIB protocol (and other QM echo protocols) for intense narrowband signal pulses ($\delta\omega_s\ll\frac{\pi\Delta_{in}}{\alpha_R(0)L}$), the equation for the pulse area of the photon echo can be used \cite{Moiseev1987, Moiseev2004Izv,Urmancheev2019,Moiseev2020}.

\fla{
 \pm\frac{\partial}{\partial z} \theta_e(z) 
&= \frac{\alpha_R(0)}{2}[2P_e(0,z)\cos^2{\frac{\theta_e(z)}{2}}
\nonumber
\\
&+W_e(0,z)\sin{\theta_e(z)}],
\label{theta}
}
where signs "$+$" and "$-$" correspond to the forward and backward geometry of the echo signal emission,
$P_e(0,z)$ is the amplitude of the phasing coherence component at a resonant frequency. 
For example, for a two-pulse photon echo $P_e(0,z)=\Gamma(t_e)\sin{\theta_1(z)}\sin^2{\frac{\theta_2(z)}{2}}$ and $W_e(0,z)$ is a component of the atomic inversion  which slowly changes from frequency detuning, taken at the central frequency. 
For the two-pulse (primary) echo $W_e(0,z)=-\cos{\theta_1(z)}\cos{\theta_2(z)}$ \cite{Moiseev2020}.
Eq.\eqref{theta} describes the pulse area of the generated echo signal in the two-level medium. 
Assuming $P_e(0,z)=0$, and $W_e(0,z)=-1$, the Eq.\eqref{theta} reduces to the well-known McCall-Hahn area theorem \cite{McCall69}.

Everywhere below we focus on the case of plane waves of signal radiation and controlling laser pulses.
The well-known solution for the pulse area of the signal (first) pulse \cite{McCall69} $\theta_s(z)=2\arctan\{ \tan\left(\frac{\theta_s(0)}{2}\right)e^{-\alpha_R(0)L/2} \}$.
By applying two laser $\pi$-pulses at an adjacent atomic transition  and inverting the frequency offsets of atoms, we obtain the following expressions for phasing polarization $P_{crib}(0,z)=\Gamma(t_e)\sin{\theta_s(z)}$ and atomic inversion $W_{crib}(0,z)=-\cos{\theta_s(z)}$ corresponding to the formation of photon echo emission in the backward CRIB protocol. 
Analytical solution of Eq. \eqref{theta} for the pulse area of echo signal in the medium  $\theta_e(0\leq z\leq L)$ is:
\fla{
\tan\left(\frac{\theta_e(z)}{2}\right)= &\Gamma(t_e)\tan\left(\frac{\theta_s(0)}{2}\right)\cdot
\nonumber
\\
&\frac{e^{-\alpha_R(0)z/2}[1-e^{-\alpha_R(0)(L-z/2)}]}{1+\tan^2\left(\frac{\theta_s(0)}{2}\right)e^{-\alpha_R(0)L}}. 
\label{theta_CRIB}
}
Below we will focus on the solution \eqref{theta_CRIB} describing the photon echo at the medium output  ($\theta_e(z=0)$):

%\fla{
%\tan\left(\frac{\theta_e(0)}{2}\right)= &\Gamma(t_e)\tan\left(\frac{\theta_s(0)}{2}\right)\cdot
%\nonumber
%\\
%&\frac{1-e^{-\alpha_R(0)L}}{1+\tan^2\left(\frac{\theta_s(0)}{2}\right)e^{-\alpha_R(0)L}}. 
%\label{eq-7}
%}

\fla{
\theta_e(0)= & 2\arctan\big\{
\Gamma(t_e)\tan\left(\frac{\theta_s(0)}{2}\right)\cdot
\nonumber
\\
&\frac{1-e^{-\alpha_R(0)L}}{1+\tan^2\left(\frac{\theta_s(0)}{2}\right)e^{-\alpha_R(0)L}} \big\}.
\label{eq-7}
}

\begin{figure}
%\begin{center}\vspace{1cm}
%\includegraphics[width=1\linewidth]{Fig5_theta_B.png}
%\includegraphics[width=1.0\linewidth]{Fig5_theta_B.eps}
\includegraphics[width=1.0\linewidth]{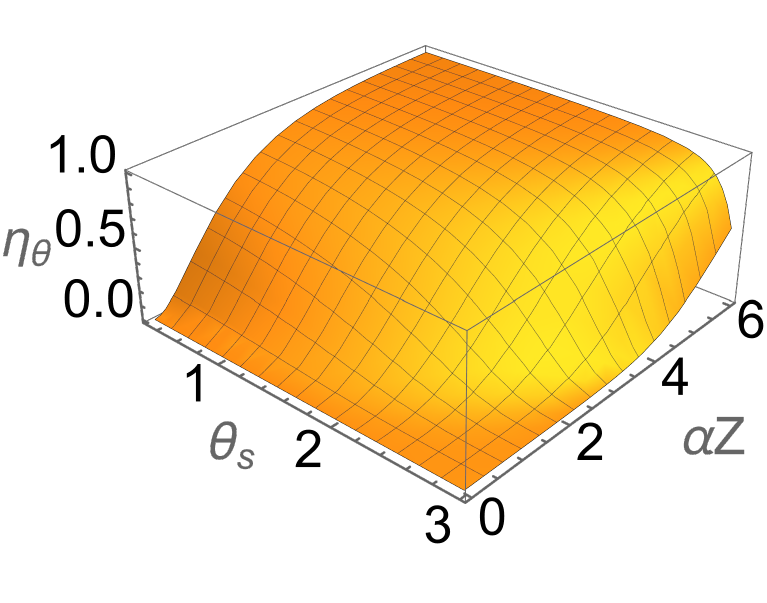}
\caption{ Efficiency of the backward CRIB-protocol measured by $\eta_{\theta}(\theta_e(z))=|\theta_e(z)/\theta_s(0)|^2$ as a function of the optical depth $\alpha Z$  and pulse area $\theta_s(0)$ of the signal pulse.}
\label{CRIB-eff backward}
\end{figure}

The solution \eqref{eq-7}  shown in the Fig. \ref{CRIB-eff backward} generalizes Eq.\eqref{eq-3} to an arbitrary pulse area $\theta_s(0)$ of the signal pulse.
%The solution \eqref{eq-7} presented in the Fig.\ref{CRIB-eff backward} generalizes \eqref{eq-3} to the  arbitrary pulse area $\theta_s (0)$ of the signal pulse.
As can be seen in the Fig.\ref{CRIB-eff backward}, increasing the pulse area of the input signal pulse   $\theta_s (0)$ close to $\pi$ leads to additional  requirement for the optical depth:  $\alpha_R(0)L\gg1$ and $\alpha_R(0)L\gg \ln\{\tan^2\left(\frac{\theta_e(0)}{2}\right)\}$ to ensure sufficiently high efficiency.
%As can be seen in the Fig.\ref{CRIB-eff backward}, increasing the pulse area of the input signal pulse makes it difficult to achieve high efficiency.
%Namely, an increase of  $\theta_s (0)$ close to $\pi$ leads to additional  requirement for the optical depth:  $\alpha_R(0)L\gg1$ and $\alpha_R(0)L\gg \ln\{\tan^2\left(\frac{\theta_e(0)}{2}\right)\}$ to ensure sufficiently high efficiency.
\noindent
The last inequality expresses the manifestation of nonlinear interaction of light with atoms (since $ \ln\{\tan^2\left(\frac{\theta_e(0)}{2}\right)\}\gg1$ when $\theta_s (0)\rightarrow \pi$).
Two limiting cases are interesting:

1) $\alpha_R(0)L\ll1$: 

$\theta_e(0)=\Gamma(t_e)\sin{\theta_s(0)}\alpha_R(0)L$, which shows that an echo signal is emitted by the atomic polarization ($\sim \sin{\theta_s(0)}$) excited by a signal pulse. 
This solution describes typical modes of the photon echo emission in the optically thin media
\cite{Hahn1950, Kopvillem1963, Kurnit1964}.

2)  $\alpha_R(0)L\gg1$:

$\theta_e(0)=2\arctan\{\Gamma(t_e)\tan\left(\frac{\theta_s(0)}{2}\right)\}$.
The solution shows a complete reconstruction of an arbitrary input signal pulse area $\theta_e(0)=\theta_s(0)$ at $\Gamma(t_e)\cong 1$.

Thus, the solution \eqref{theta_CRIB}, \eqref{eq-7} makes it possible to understand the effect of phase relaxation of atoms and the optical depth of the medium on the reconstruction of an echo signal in conditions of nonlinear resonant interaction with atoms.
It should be noted that the study of these effects cannot be carried out using the well-known method of the inverse scattering problem \cite{Ablowitz1974}, which is aimed at finding analytical solutions for light pulses (solitons) arising in the depth of the medium without taking into account the influence of phase relaxation of atoms \cite{Maimistov1990}.
%Studies of these effects cannot be achieved using the well-known inverse scattering method  \cite{Ablowitz1974} used in finding analytical solutions for light pulses (solitons) in the depth of the medium neglecting the influence of phase relaxation of atoms \cite{Maimistov1990}.

Solution \eqref{eq-7} reproduces the reversibility of the nonlinear Maxwell-Bloch equations in the CRIB protocol \cite{Kraus2006},
demonstrating that the photon echo area theorem \eqref{theta} plays the role of a time-reversed version of the McCall-Hahn area theorem \cite{McCall69}.
%Given the reversibility of the nonlinear Maxwell-Bloch  equations  in the CRIB protocol \cite{Kraus2006}, the solution  \eqref{eq-7} indicates that the photon echo area theorem \eqref{theta} plays the role of a time-reversed version of the McCall-Hahn area theorem \cite{McCall69}.
Thus the photon echo area theorem \cite{Moiseev1987,Moiseev2020} provides important tool for analytical study of photon echo at the conditions of nonlinear light-atoms interaction and can be useful for studies of different echo signals \cite{Moiseev2020,Moiseev_Urmancheev2022} and QM echo-protocols, as it will be used below.

Substituting the same values of phasing polarization $P_{crib}(0,z)$ and atomic inversion $W_{crib}(0,z)$ in Eq. \eqref{theta}, but taking into account the propagation of the photon echo in the forward direction,  we find the following solution for its pulse area $\theta_e(z)$ in the medium:

\fla{
\tan\left(\frac{\theta_e(z)}{2}\right)= &\Gamma(t_e)\tan\left(\frac{\theta_s(0)}{2}\right)\cdot
\nonumber
\\
&\frac{\alpha_R(0)ze^{-\alpha_R(0)z/2}}{1+\tan^2\left(\frac{\theta_s(0)}{2}\right)e^{-\alpha_R(0)z}}. 
\label{CRIB-forw}
}

\begin{figure}
%\begin{center}\vspace{1cm}
%\includegraphics[width=1\linewidth]{Fig6a_e_tan_F.png}
\includegraphics[width=1.0\linewidth]{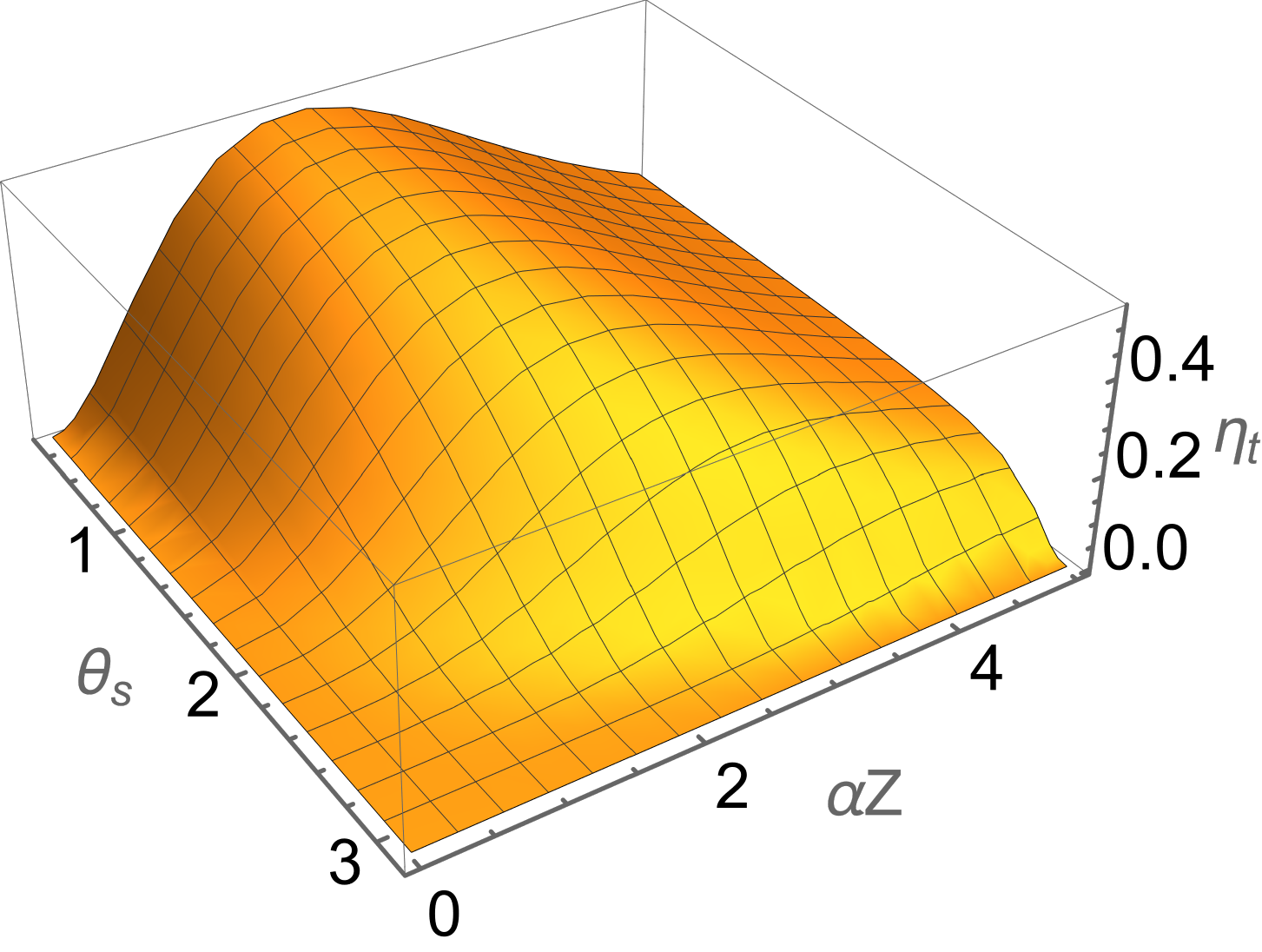}
\caption{Efficiency of the  forward CRIB-protocol measured by $\eta_t(\theta_e(z))=|\frac{\tan\left(\theta_e(z)/{2}\right)}{\tan\left(\theta_s(0)/{2}\right)}|^2$ as a function of  $\alpha_R(0) Z$  and $\theta_s(0)$ of the signal pulse.}
\label{CRIB-forward eff tan}
\end{figure}

\noindent
For weak input signal pulse $\theta_s(0)\ll1$ in \eqref{CRIB-forw}, we obtain $\theta_e(z)=\Gamma_R(t_e)\theta_s(0)\alpha_R(0)ze^{-\alpha_R(0)z/2}$ corresponding to the above linear solutions for the photon echo of weak spectrally narrow signal pulse \eqref{eq-5}.
From the Eq.\eqref{CRIB-forw}, we found that with increasing $\theta_s(0)$ the efficiency reaches its maximum at a larger optical depth $\alpha_R(0)L=2/|\cos{\theta_s(L)}|\geq 2$).
If we evaluate the overall performance behavior of the protocol in Eq.\eqref{CRIB-forw} using a seemingly acceptable ratio $\eta_t(\theta_e(z))=|\frac{\tan\left(\theta_e(z)/{2}\right)}{\tan\left(\theta_s(0)/{2}\right)}|^2$, we get fast decrease of $\eta_t(\theta_e(z))$ with increasing $\theta_s(0)$ and  $\alpha_R(0)L$ (see Fig.\ref{CRIB-forward eff tan}).

\begin{figure}
%\begin{center}\vspace{1cm}
%\includegraphics[width=1\linewidth]{Fig7b_e_theta_F.png}
\includegraphics[width=1.0\linewidth]{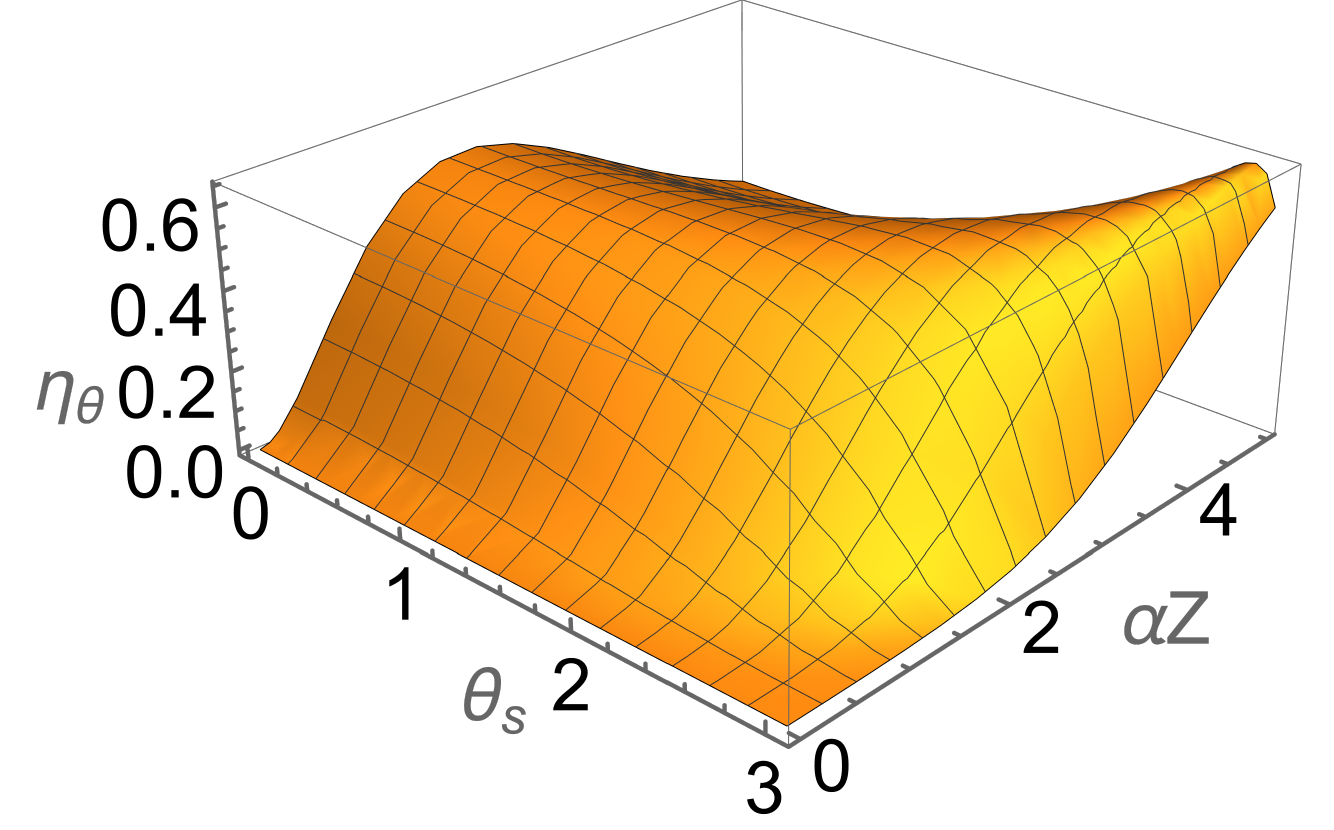}

\includegraphics[width=1.0\linewidth]{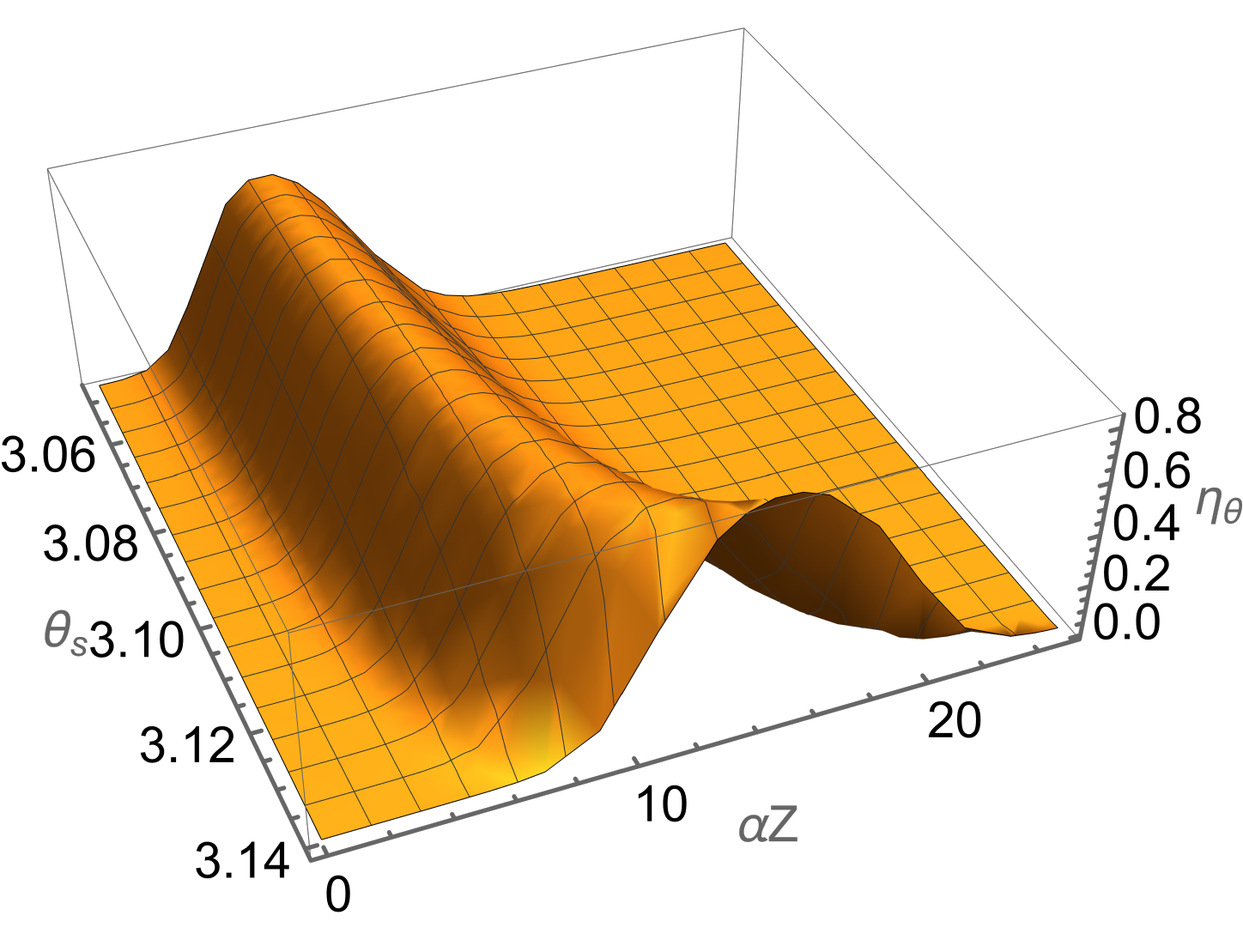}
\caption{ Efficiency of the  forward CRIB-protocol measured by $\eta_{\theta}(\theta_e(z))=|\theta_e(z)/\theta_s(0)|^2$ as a function of  $\alpha_R(0) Z$  and $\theta_s(0)$, where top ($\theta_s(0):0.0\pi-0.97\pi$), bottom ($\theta_s(0):0.97\pi-0.9995\pi$).}
\label{CRIB-forward eff teta}
\end{figure}

It is more correct to use a measure $\eta_{\theta}(\theta_e(z))=|\theta_e(z)/\theta_s(0)|^2$, which is confirmed in experiments with light pulses, the spectral width of which is much smaller than the inhomogeneous broadening of the resonant transition \cite{Minnegaliev2021,Moiseev2023}.
Under these conditions, the time profile of photon echo signals can reproduce the time profile of signal pulses even under conditions of strong nonlinear interaction, but not at too large  optical depth of the resonant medium, where optical solitons have not yet had time to form.
Using the efficiency measure $\eta_{\theta}(\theta_e(z))$, we  find that the efficiency can increase asymptotically approaches unity at $\theta_s(0)\rightarrow\pi$ (see Fig.\ref{CRIB-forward eff teta}), reflecting the nonlinear nature of the self-induced transparency effect near the bifurcation point $\theta=\pi$, but then falling to zero over long distances (see Fig.\ref{CRIB-forward eff teta}).

Thus, the enhancement of the nonlinear interaction of the signal pulse with a two-level medium leads both to an increase in the optimal optical depth and to maximum protocol efficiency.
The counter intuitive enhancement of the efficiency is obviously associated with  an increase of the penetration depth of an intense light pulse into the medium caused by the effect of self-induced transparency \cite{McCall69}.
However, these efficiency measures based on the estimation of the echo pulse area,  require further clarification when switching to using on an energy efficiency measure. 

The study of quantum properties in the storage of intense signal pulses is already gaining interest for optical quantum communications \cite{Vinh2021} and it can be carried out using the capabilities of the inverse scattering method  \cite{Ablowitz1974,Maimistov1990} and its quantum generalization \cite{Rupasov1982}. 
This, however, will require its further development in obtaining non-soliton solutions.

The discussed properties of CRIB protocol are basically repeated in studied below AFC- and ROSE-protocols, with the advent of some new features and practically significant advantages.
However, the departure from the exact reversibility of the light-atoms equations of motion in some cases lead to the additional problems in effective implementation of these protocols.
We discuss some of them in the next sections.

The Gradient Echo Memory (GEM) protocol  has been proposed  as a way to implement the CRIB protocol \cite{Alexander2006}, where the CRIB-procedure ($\Delta_j\rightarrow-\Delta_j$) is provided using a switchable electric field gradient along the propagation of a signal light pulse ($\Delta=\chi z$) causing a linear Stark effect of narrow homogeneously broadened optical transition in REI ensemble, where $\chi L$  is the spectral width of the GEM cell that satisfies the condition $\chi L>\delta\omega_s$.

The GEM protocol has been demonstrated in solids and atomic gases \cite{Lauritzen2010,Hedges2010,Hosseini2011}. 
In \cite{Hedges2010}, an efficiency of 69\% was achieved on a Pr$^{3+}$:Y$_2$SiO$_5$ crystal. 
Due to the application of the Zeeman effect with a switchable magnetic field gradient in \cite{Hosseini2011,Hosseini2009,Cho2016}, using GEM protocol with Raman transition ($\Lambda$-scheme) of rubidium in a vapor gas, the efficiency of 87\% was achieved for quantum states of light, which remains the best among optical quantum memory protocols.

\begin{figure}
%\begin{center}\vspace{1cm}
%\includegraphics[width=1.0\linewidth]{Fig8-GEM.png}
\includegraphics[width=1.0\linewidth]{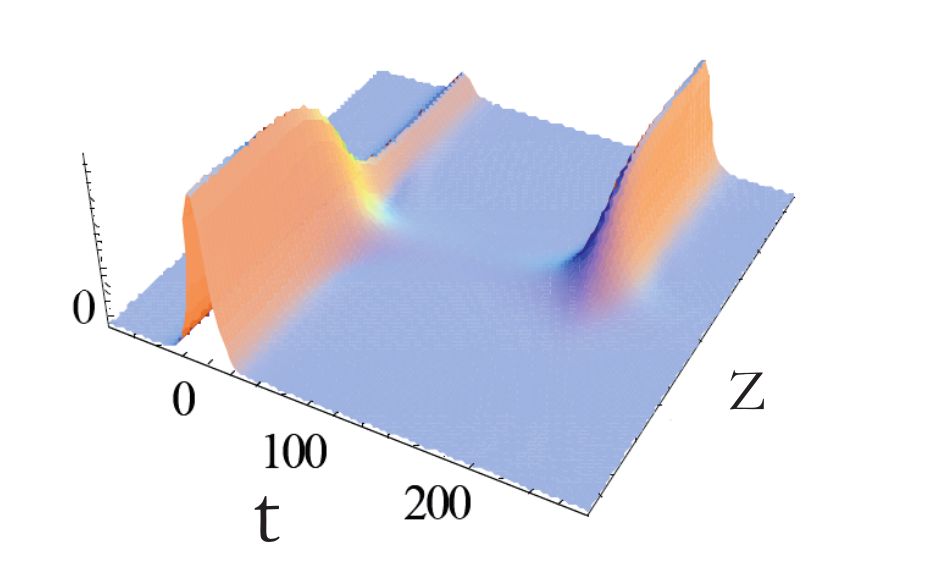}
\caption{An efficient forward scheme of GEM protocol:
storage and retrieval of a light pulse with an pulse area of $18^o$, the absorption coefficient $\varkappa_{eff}=1$, the gradient sign is changed at $t=100$ (numerical results of \cite{Saidasheva2006}).}
\label{GEM-forward}
\end{figure}

The GEM protocol coincides with the CRIB protocol when signal is retrieved in the backward direction, respectively, while having full temporal reversibility of the Maxwell-Bloch equations corresponding to the stage of absorption of the signal pulse and the stage of its retrieval in the photon echo.
However, as it was firstly shown numerically   \cite{Saidasheva2006}  in the GEM protocol it is possible to achieve efficiency close to 100\%, not only in the backward but also in the forward direction of the photon echo emission (see Fig.\ref{GEM-forward}).
The efficiency of GEM-protocol with forward and backward echo emission is described by the same equation \eqref{eq-3} where $\alpha_R(0)L$ is replaced by an effective optical depth $\varkappa_{eff}=2\beta/\chi$ \cite{Moiseev2008,Hosseini2011}. 
This useful property arises due to the fact that resonant atoms are always located at one point in the medium, which does not lead to reabsorption of echo signals and gives great advantages in using this protocol.
In particular, the retrieval of the signal field in the forward direction makes GEM protocol very  attractive  for use in quantum optical holography.

At the same time, due to the violation of temporal reversibility, the echo signal emitted in the forward direction acquires additional nonlinear phase modulation factor $e^{i\phi_n(t)}$ \cite{Moiseev2008}: 
\fla{
\tilde{a}_e(t)\sim \Gamma(t_e )\tilde{a}_s(t_e-t) e^{i\phi_n(t)},
\label{phase-modulation}
}
where the phase  
$\phi_n(t)=\varkappa_{eff}Ln\{1+\frac{t-t_e}{t_1+t_m}\}$, 
$t_m=\varkappa_{eff}/(\chi L)$, $t_e$ 
is the time moment of photon echo emission.
Phase chirping caused by the nonlinear behavior of the phase $\phi_n(t)$ is weakened if the interaction time $t_1$ of a signal pulse with the atoms is longer than the pulse duration ($ t_1\gg \delta t_s\sqrt{\varkappa_{eff}/\pi}$, for $t_m\ll t_1$).
Therefore, after the signal has entered the medium, it is necessary to wait a certain time $t_1$ before turning off the external gradient. 
This requirement arises because in the space region of the exact resonance, the signal pulse experiences a significant slowdown in its propagation  due to the manifestation of strong spectral dispersion.

Characterizing the specific properties of GEM-protocol, it is also important to note that the signal pulse is stored in the spatial shape of the excited atomic coherence. 
Unlike the slow-light QM protocol \cite{Lukin2003}, the spatial shape  is determined by the shape of the signal pulse spectrum.
Accordingly, unlike slow-light QM, a large number of signal pulses can be stored in the same area of the QM cell space, so the number of which is only limited by the ratio of inhomogeneous broadening to the homogeneous linewidth of the resonant transition ($\chi L/\gamma\gg 1$).
At the same time, the specific nature of the inhomogeneous broadening of the resonant transition of the GEM-protocol does not allow the application of the pulse area theorem in the study of general nonlinear patterns of interaction of light pulses with atoms, the understanding of which remains an unsolved theoretical task.

The ability to control the magnitude and sign of the gradient of the external electric (or magnetic) field gives researchers additional opportunities to control the operation of GEM-protocol.
These are particularly relevant to the GEM protocol on the Raman transitions. 
The following can be noted:

1) Typically, the CRIB/GEM protocol is characterized by obtaining a photon echo with a time-reversed shape compared to the time shape of the signal pulse.
Using Raman transitions in GEM protocol, it is possible to restore the photon echo, which will have the temporal shape of a signal pulse.  
To do this, after absorption of a signal pulse we first reverse the frequency detuning of atoms ($\Delta \rightarrow -\Delta$) without using a control laser pulse, ensuring complete dephasing of the spin coherence of atoms with a sufficiently long time of evolution of spin coherence.  
Then change the frequency offsets again ($-\Delta \rightarrow \Delta$) in the presence of the control laser pulse will lead to the photon echo emission with a time profile of the signal pulse.

2) The possibility of reducing the inhomogeneous broadening of the optical transition by switching off the gradient of the applied electric (magnetic) field makes it possible to realize enhanced off-resonant interaction of the quantum coherence stored in the medium with probe quantum light fields. 
Such interactions are used to provide nonlinear Kerr interactions of single-photon wave packets in order to deterministically implement the conditional cross-phase shift gates \cite{Lukin2000,Wang2006,Marzlin2010,Scherer2012,Bienias_2020}.
Recently demonstrated GEM protocol  \cite{Leung_2022} appears to be a promising alternative to the slow-light protocol \cite{Lukin2000,Feizpour2015} in the deterministic implementation of two-photon qubit gates, which can be free from the limitations inherent in the use of nonlinear interaction of two propagating photon wave packets detailed in \cite{Shapiro2006,Banacloche2010}.

3) It is also worth noting that by changing the magnitude of the gradient of the electric (magnetic) fields, the duration of the photon echo signal can be manipulated while maintaining high efficiency \cite{Hosseini2009,Moiseev2010}, which is impossible when using a transverse inhomogeneous broadening  to the propagation of the signal field (that is, for the usual CRIB protocol \cite{Moiseev2001}). 

4)Another promising area of Raman GEM protocol development is the creation of a quantum light generation using reservoir engineering with a help of four-wave interaction \cite{MoiseevES2020}.

5) The record efficiency achieved in the  GEM protocol on the Raman transition in atomic gas stimulated work on the implementation of QM echo protocols in crystals with REIs \cite{Moiseev2013PRA,Kalachev2013}.
The solution to this problem is possible by using a resonator schemes with crystals having a sufficiently narrow inhomogeneous broadening of the optical transitions.
The search for such crystals continues \cite{2015-Goldner-Handbook,Popova2019,2023-Gerasimov-OandS,Gerasimov2024}.

6) It is worth noting that the inverted time form of the photon echo signal in CRIB/GEM protocol allows the implementation of a NOT gate plus phase gate with a signal pulse prepared in the time-bin state of the photonic qubit
\cite{MoiseevTittel2010}.
The implementation of CNOT gate is a big challenge for further development of the CRIB/GEM  protocol. 
Among the possible ways to solve it, one can try to use a two-resonator scheme of time-bit quantum RAM scheme \cite{Moiseev2016}.

A significant limitation of the CRIB/GEM protocol is the difficulty of achieving the required high optical depth of the resonance transition in a wide spectral region. 
This problem can be solved to some extent by additional using methods that allow significantly increasing the constant of photon-atom interaction. 
For example, this is possible when implementing this protocol in a high-Q resonator \cite{EMoiseev2021}, by using surface plasmon polaritons \cite{Tame2013,Mois2010} instead of flying light fields and applying other nanooptics methods\cite{Novotny2012}.

\section{AFC protocol}
The Atomic Frequency Comb (AFC) protocol uses an ensemble of atoms with equidistantly spaced narrow resonant lines $\Delta_n=n\Delta_{afc}$ with $\Delta_{afc}$ is the distance between the lines (n=0,±1,±2,...,). 
Such inhomogeneous broadening was first proposed for photon echo in 1985 and demonstrated on a system of nonlinear oscillators in \cite{Dubetskii1985,Dubetskii1986}.

After excitation by a resonant pulse at time $t=0$, the n-th group of oscillators goes into free oscillation, acquiring a phase $\sim e^{-in\Delta_{afc} t}$. When the phases of all oscillators coincide again at time $t_n=2\pi n/ \Delta_{afc}$, the echo signal is emitted.
In 2008, this variant of inhomogeneous broadening was proposed  for echo-based QM in \cite{DeRiedmatten2008}.
During the interaction time of a short light pulse with AFC structure $\delta t_s\ll\Delta^{-1}$ of the inhomogeneous line broadening is characterized by an averaged (effective) optical density 
$\alpha_{afc} (\omega)=\frac{\Upsilon}{\Delta_{afc}}\alpha (\omega)$, where $\Upsilon$ is the spectral width of a single peak, $\alpha (\omega)$  is determined from \eqref{eq-2},  $f=\Delta_{afc}/\Upsilon$ - is a finesse  of AFC structure.
Due to this property, the MB equations can be solved independently for each non-overlapping input light pulse at absorption and echo emission stages.
A solution for the Fourier component of an echo signal 
$\hat{a}_e(Z,t)=\frac{1}{\sqrt{2\pi}}\int d\omega \tilde{a}_e(Z,\omega)e^{-i\omega(t-t_e-Z/v_g)}$
emitted in parallel to the signal pulse after its absorption in the medium has the form:

\fla{
\tilde{a}_e(Z,\omega)=&
\Gamma_{afc}(t_e )\tilde{a}_s(0,\omega)\cdot
\nonumber
\\
&\alpha_{R,afc} (\omega)Z 
e^{-\alpha_{afc}(\omega)Z/2},
\label{eq-9}
}
where  $\alpha_{R,afc} (\omega)=\alpha_{R} (\omega)/f$ 
is an effective absorption coefficient, $\Gamma_{afc}(t_e=1/\Delta_{afc} )=e^{-7/2f^2}$ \cite{Afzelius2009},  which requires the creation of a high finesse  AFC structure with $\Delta_{afc}/\Upsilon>20$ to get $\Gamma_{afc}(1/\Delta_{afc} )\cong 1$. 
Since $\alpha_{afc} (\omega)=\alpha_{R,afc}(\omega)(1+i\frac{\omega}{\Delta_{in}})$ 
(for Lorentzian   inhomogeneous broadening), 
and  $\alpha (\omega)=\alpha_{R}(\omega)(1+i\frac{\omega}{\Delta_{in}})$
the echo signal in \eqref{eq-9} 
acquires phase modulation (see in comparison \eqref{eq-3}), which obviously distorts the time profile of the broadband echo signal, but not affects the spectral efficiency of the AFC protocol
$\eta_L(\omega,Z)\sim [\alpha_{R,afc}(\omega)Z]^2\exp\{-\alpha_{R,afc}(\omega)Z\}$.

The maximum of $\eta(\omega,Z)$ remains equals to 52\% at $\alpha_{R,afc}(0)L=2$ for narrowband signal pulse ($\delta\omega_s\ll\Delta_{in}$) and $\Gamma_{afc}(1/\Delta_{afc} )\cong 1$, that is similar to the CRIB protocols.
However, the echo pulse  $\hat{a}_e(Z,t)$ of \eqref{eq-9} reproduces the temporal shape of the signal pulse that indicates a violation of the temporal reversibility in the AFC protocol.
Taking into account the spectral dispersion in $\alpha_{afc} (\omega)$, we get that narrowband echo signal propagates with modified group velocity  
$\tilde{v}_g=v_g\frac{1}{1-\frac{v_g\alpha_{R,afc}(0)}{2\Delta_{in}}}$, which leads to an additional (negative) time delay  in the emission of the echo signal   $\delta t_e=-\frac{\alpha_{R,afc}(0)L}{2\Delta_{in}}$ caused by the interaction of the light pulses with resonant atoms.
For the maximum efficiency (i.e. at $\alpha_{R,afc}(0)L=2$) we have the same time delay $\delta t_e=-\Delta_{in}^{-1}$.

Now we consider AFC protocol in a 3-level scheme with echo emission  in the backward direction. 
This is implemented similarly to the CRIB protocol (\cite{Moiseev2001} and see above) by applying two antiparallel laser $\pi$ pulses at an adjacent atomic transition after signal pulse absorption (see backward geometry in Fig.\ref{CRIB}).
After similar calculations, we get:

\fla{
\tilde{a}_e(0,\omega)=&
\Gamma_{afc}(t_e )\tilde{a}_s(0,\omega)\cdot
\nonumber
\\
& 
\frac{\alpha_R(\omega)}{\alpha(\omega)}
\{1-e^{-\alpha_{afc}(\omega)L}\}.
\label{eq-10}
}
\noindent
At low optical depth ($\alpha_{R,afc }L\ll1$) in Eq. \eqref{eq-10},  we have
$\tilde{a}_e(0,\omega)=
\Gamma_{afc}(t_e )\tilde{a}_s(0,\omega)\alpha_{R,afc}L$, 
this indicates no influence of spectral dispersion to the  echo pulse parameters at low efficiency, which is observed  experimentally.
At a high optical depth ($\alpha_{R,afc }L\gg1$),  we get by assuming  the  Lorentzian  inhomogeneous broadening:

\fla{
\tilde{a}_e(0,\omega)=&
\Gamma_{afc}(t_e )\tilde{a}_s(0,\omega)
\frac{1-i\frac{\omega}{\Delta_{in}}}{1+\frac{\omega^2}{\Delta_{in}^2}}.
\label{eq-11}
}

It is seen in Eq.\eqref{eq-11} that the spectral efficiency  $\eta_L(\omega)$ of the AFC protocol decreases with an increase in the spectral width of the signal pulse $\eta_L(\omega) = \left(1+\frac{\omega^2}{\Delta_{in}^2}\right)^{-1}$, which leads to a narrowing of the spectrum of the restored signal

\fla{
|\tilde{a}_e(0,\omega)|^2\sim
\frac{|\tilde{a}_s(0,\omega)|^2}{(1+\frac{\omega^2}{\Delta_{in}^2})}.
\label{spectrum}
}

By taking into account that $(1-i\frac{\omega}{\Delta_{in}})\cong \exp\{-i\frac{\omega}{\Delta_{in}}\}$ for   narrowband signal pulse ($\delta\omega_s\ll\Delta_{in}$), we get from Eqs. \eqref{eq-10}, \eqref{eq-11} that the spectral dispersion only leads  to an additional time delay in the emission of the echo signal $\delta t_e=-\Delta_{in}^{-1}$.
However in the case of broadband photon echo signal ($\delta\omega_s\sim\Delta_{in}$), the spectral dispersion  causes too large phase distortions that reduce the efficiency and fidelity of the signal pulse retrieval \cite{Moiseev2012}.
The sensitivity of a high-performance AFC protocol to the spectral dispersion distinguishes it from the CRIB protocol, which does not acquire additional phase distortions.
The need to suppress the spectral dispersion caused by interaction with resonant atoms becomes important already when an efficiency of more than 60\% is realized \cite{Moiseev2012}.
The increasing the operation spectral range $\Delta_{qm}$ while maintaining high efficiency and fidelity when using AFC protocol requires an additional suppression of the spectral dispersion.

The existing experiments on AFC protocol in crystals with RE ions are being implemented by creating the AFC structure of a certain spectral interval inside an inhomogeneously broadened resonant transition \cite{Guo2023,Zhou2023}.
Taking into account such a spectral  design of AFC structure, it is possible to propose an optimal way to create such a structure in which spectral dispersion will be maximally suppressed in a given operating spectral range of AFC protocol \cite{Arslanov2017,Arslanov2019}.
The approach \cite{Arslanov2017,Arslanov2019} is based on the creation of AFC structure with optimal spectral parameters inside a inhomogeneously broadened line for the  given operating spectral range $\Delta_{qm}$ inside this AFC-structure and necessary protocol efficiency \cite{Arslanov2017}. 

\begin{figure}
%\begin{center}\vspace{1cm}
%\includegraphics[width=1.0\linewidth]{Fig9-AFC-mod-1.png}
\includegraphics[width=1.0\linewidth]{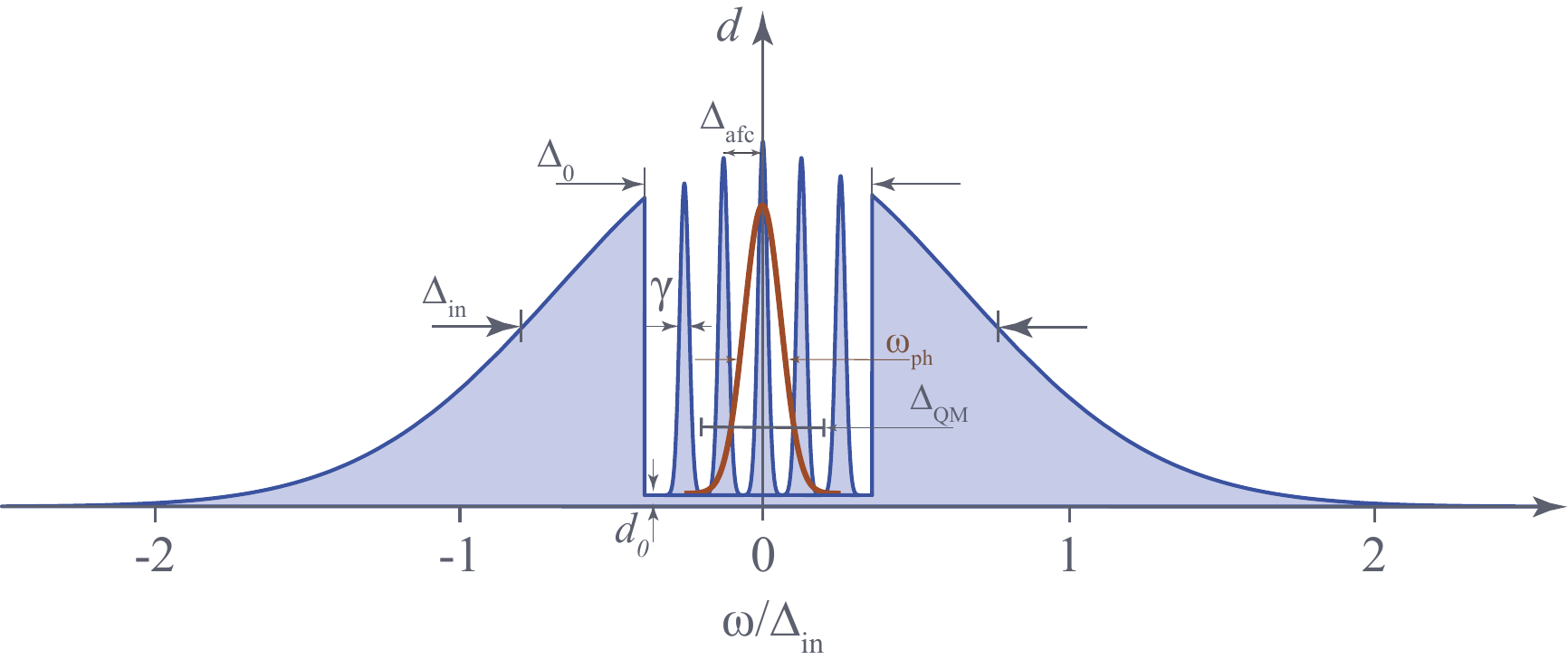}
\caption{
Spectral design of atomic frequency comb  created within spectrum $\Delta_0=1.25\Delta_{in}$ of inhomogeneously broadened atomic transition with finesse $f=\Delta_{afc}/\Upsilon=10$, providing efficiency 90\% within operating spectral range $\Delta_{QM}=0.9\Delta_{in}$, optical depth $\alpha_R(0)L=80$  \cite{Arslanov2019}.}
\label{AFC-struct}
\end{figure}

\begin{figure}
%\begin{center}\vspace{1cm}
%\includegraphics[width=1.05\linewidth]{Fig10_AFC_1.png}
\includegraphics[width=1.0\linewidth]{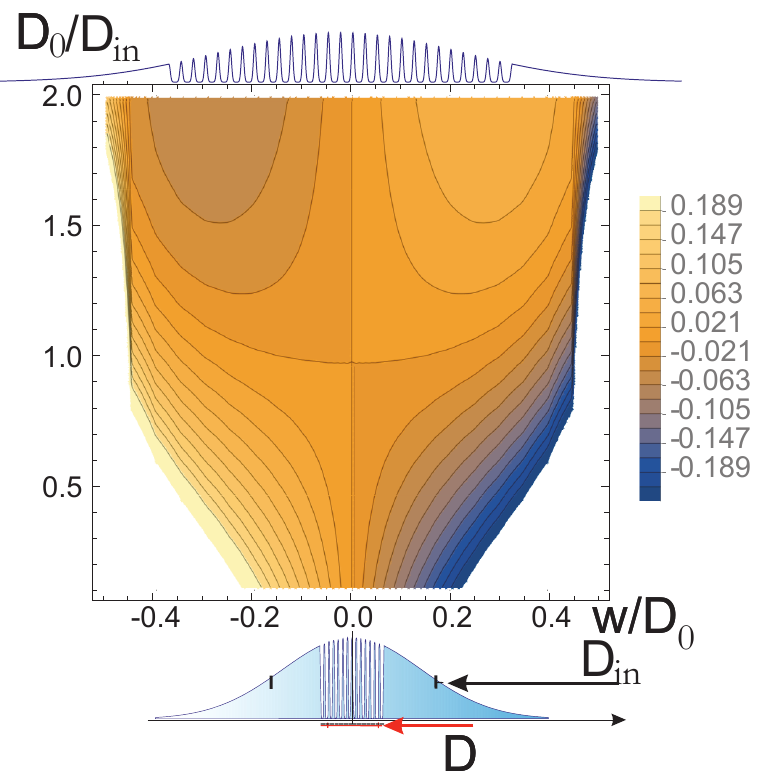}
\caption{
The the contour map image shows   $\chi'(\omega)$  within AFC structures $\omega/\Delta_0$ for the various AFC structure witdh $\Delta_0$ is changed within inhomogeneously broadening  $\Delta_{0}/\Delta_{in}$. 
Due to the spectral dispersion, there is some area where the opposite effects of the frequency response of  AFC structure and the side "wings"  take place causing the creation of a plateau with a slowly changing line.}
\label{AFC-1-10}
\end{figure}

\begin{figure}
%\begin{center}\vspace{1cm}
%\includegraphics[width=1.0\linewidth]{Fig11_AFC_3.png}
\includegraphics[width=1.0\linewidth]{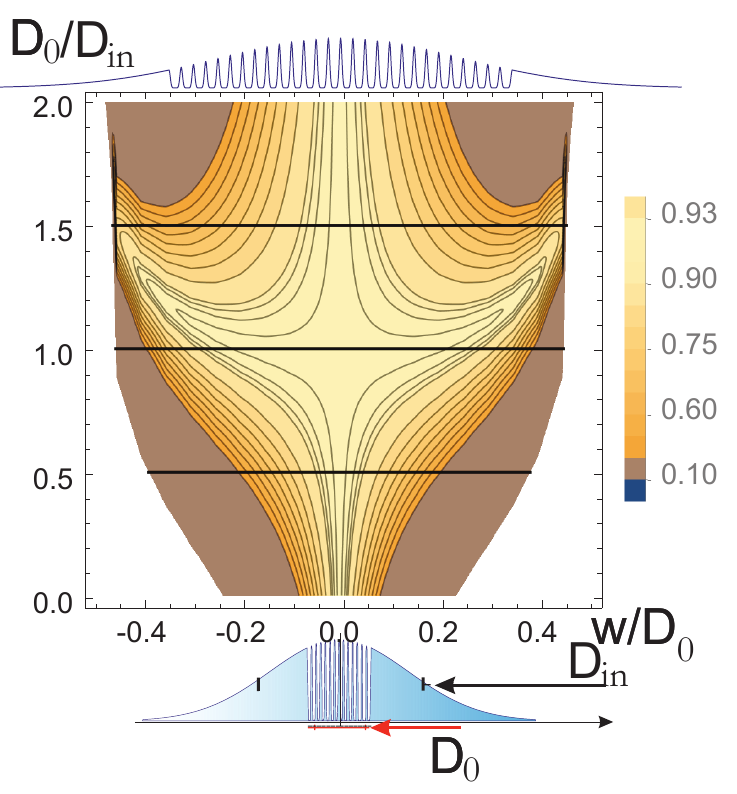}
\caption{The the contour map image shows spectral efficiency $\eta_{afc}(\omega)=|\Gamma_{afc}(t_e)* \mathcal{D}(\omega)|^2$, within AFC structures $\omega/\Delta_0$ for the various width $\Delta_0$ of AFC structure is changed within inhomogeneously  broadening  $\Delta_{0}/\Delta_{in}$.}
\label{AFC-3-11} 
\end{figure}

Full consideration of the effect of the spectral dispersion of atoms when creating an AFC structure in a limited frequency range $\Delta_0$ (see Figs.\ref{AFC-struct}-\ref{AFC-3-11}) of an inhomogeneously broadened transition includes two terms.
A detailed analysis of their influence is the subject of a separate study, taking into account the real parameters of the optical transition \cite{2024-Arslanov-inprogress}. 
One of them is due to the frequency dispersion of the AFC structure itself, and the other is caused by the influence of the remaining two spectral regions (side "wings"). 
Atoms whose frequencies belong to the side wings of the resonant transition change the frequency dispersion without practically affecting the resonant absorption coefficient, leading to a specific change in $\alpha_{afc} (\omega)$
to a form  $i \alpha_{R, afc} (\omega) \chi (\omega)$ \cite{Moiseev2012}, where
\fla{
\chi(\omega)=&\chi'(\omega)+i\chi"(\omega)
\nonumber
\\
=&\chi_0(\omega)+\chi'_1(\omega)+\chi'_2(\omega).
\label{ref_index}
}
\noindent
The  first term in  $\chi(\omega)$ is caused by the  from AFC structure 
$\chi_0(\omega)=\chi'_0(\omega)+i\chi''_0(\omega)$,
where by taking into account Gaussian shape of the inhomogeneous broadening
\fla{
\chi'_0 (\omega)=&-\frac{1}{f} \exp{[-\zeta \omega^2/\Delta_{in}^2]} Erfi(-\sqrt{\zeta} \omega/\Delta_{in}),
 \nonumber
 \\
\chi''_0 (\omega)=&-\frac{1}{f} \exp{[-\zeta \omega^2/\Delta_{in}^2]},
}
\noindent
where $\zeta = 4Ln2$,  $\alpha_{R, afc} (\omega) \chi_0''(\omega)$ describes the average value of the absorption coefficient within the AFC-structure. 

From the left and right "wings" of the inhomogeneously  broadened line we have  dispersion terms  
\fla{
\chi'_1 (\omega)= (1-\frac{1}{f})\frac{1}{\pi}\int\limits_{-\infty}^{-\Delta_0/2\Delta_{in}} \frac{\exp{[-\zeta x^2]}}{\frac{\omega}{\Delta_{in}}-x} dx,
\\
\chi'_2 (\omega)=(1-\frac{1}{f})\frac{1}{\pi}\int\limits_{\Delta_0/2\Delta_{in}}^{\infty} \frac{\exp{[-\zeta x^2]}}{\frac{\omega}{\Delta_{in}}-x} dx.
}
The amplitude of photon echo in backward AFC protocol is given by  \cite{Arslanov2017,Arslanov2019}  
\fla{
\tilde{a}_e(0,\omega)=
\Gamma_{afc}(t_e )\tilde{a}_s(0,\omega)
\mathcal{D}(\omega), 
}
where
\fla{
\mathcal{D}(\omega)=\frac{1-exp{[i\omega_0\chi (\omega) L/c]}}{1-i\chi'(\omega)/\chi''(\omega)}.
\label{eq-11_2}
}
The behavior of the dispersion $\chi'(\omega)$ depending on the spectral width $\Delta_0$  of the AFC structure is shown in the Fig.\ref{AFC-1-10}. 
It is seen that there is a plateau where $\chi'(\omega)\approx 0$ and $\frac{d}{d\omega}\chi'(\omega)\approx 0$ within spectral range $-0.25\Delta_{in}<\omega<-0.25\Delta_{in}$ for $\Delta_0/\Delta_{in}\cong 1$. 
The efficiency of AFC protocol
($\eta_L(\omega)=|\mathcal{D}(\omega)|^2$)
is shown in Fig.\ref{AFC-3-11}, where its high value ($>0.93$) exists in the same spectral range  with the used parameters. 
Moreover the spectral scheme \cite{Arslanov2019}, which provides strong suppression of spectral dispersion, allows achieving protocol efficiency of more than 95\% for broadband light pulses in shown in Fig.\ref{AFC-3-11}.

The discussed effects of spectral dispersion are due to the lack of precise temporal reversibility of the quantum memory protocol. 
Note that the strong negative effect of dispersion effects becomes significant when trying to achieve high efficiency in a wide spectral range compared  with the inhomogeneous broadening $\Delta_{in}$ of the resonant transition.  
A large influence of spectral dispersion also occurs when implementing the protocol in a resonator, the study of the specific properties of which requires special consideration \cite{EMoiseev2021}.

The AFC protocol was experimentally implemented at different wavelengths in several crystals with different REI \cite{Holzapfel2020,Stuart2021,Cruzeiro2018,Akhmedzhanov2016,Rielander2014,Akhmedzhanov2023}. 
In these studies, the AFC structure are usually created in a small spectral range $\Delta_0$ of an inhomogeneously broadened optical transition. 
The magnitude of $\Delta_0$ and finesse $f$  have not yet been chosen optimally to ensure strong suppression of spectral dispersion.
An efficiency of 62\% was achieved \cite{Duranti2023}, a storage time ($T_{st}$) of 1 hour \cite{Ma2021}, fidelity of  $>$ 99\% \cite{Zhu2022,Zhou2012} (for efficiency of 11 and 7$\%$), the storage of more than 1000 temporal light modes \cite{Businger2022,Wei2024,Bonarota2011} and a large operating spectral width 
($\Delta_{0} > 6$ GHz)  were demonstrated \cite{Saglamyurek2011,Wei2024, 2020-PRR-Tittel} but with efficiency of $<10\%$.
In these experiments, the efficiency of AFC protocol was severely limited by the small finesse. 
Increasing finesse to values $f>10$ with appropriate suppression of the negative effect of spectral dispersion will require an increase in the initial optical depth of the atomic transition $\alpha_R(0)>30$.
This will be possible when using a crystal that combines the presence of a high concentration of ions and a long optical coherence time.

So far, practically significant results have been obtained in different experiments and on different crystals.
In most of the above experiments, AFC memory acted as a passive delay line. 
 For practical applications, on-demand reading is required. 
 This can be achieved in two ways.
 Firstly, the Stark effect can be applied as it was implemented in the works \cite{Horvath2021,Craiciu2021}. 
 However, this method has a limitation on storage time associated with a finite AFC peak width ($\approx$ 100 kHz).
 Another way to implement on demand signal reading is transferring optical coherence using a $\pi$ pulse to an additional pre-emptied spin sub-level (see discussion of Eq. (10)). 
 At this point, it is also possible to use dynamic decoupling pulses to increase storage time. Such schemes were implemented in the following works \cite{Corrielli2016,Holzapfel2020,Rakonjac2021,Ortu2022,Alqedra2024,Businger2020}.

 The real way to increase the efficiency of AFC protocol is associated with its implementation in an optical resonator.
 The AFC protocol was already demonstrated in an optical resonator \cite{Sabooni2013,Jobez2014,Duranti2023}, in an integrated waveguide circuit \cite{Liu2022,Rakonjac2022,Saglamyurek2015} and in a photonic-crystal cavity \cite{Craiciu2019,Craiciu2021}. 
 The maximum efficiency $\eta$ $\approx$ 60 \%   was achieved for storage time of $\approx$ 1$\mu$s \cite{Sabooni2013,Duranti2023}. 
The results demonstrate a great practical potential of this protocol even in conditions of low efficiency.
In this case, it will be necessary to resolve the issue of increasing the working bandwidth, since it cannot become large when using high-quality resonators
\cite{EMoiseev2021}.

\section{ROSE protocol}

The use of a natural inhomogeneous broadening of atomic transition would significantly facilitate the achievement of high efficiency and a large spectral width.
In 2011, QM echo protocols were proposed using the natural inhomogeneous broadening of the atomic transition \cite{McAuslan2011,Damon2011,Moiseev2011PRA-PE, Beavan2011}, among which the protocol  using the revival of silenced  echo signal (ROSE) received the most attention \cite{Damon2011}.

The original scheme of ROSE protocol  was proposed for a system of two-level atoms.
The retrieval of the stored signal light pulse is implemented by using two subsequent rephasing   $\pi$ pulses. 
The ROSE echo signal is the result of the primary echo retrieval, the radiation of which is blocked by suppression of wave synchronism, or controlled dephasing of atomic coherence by external fields \cite{McAuslan2011,Arcangeli2016,Ham2017}.
Thus, the ROSE protocol turns out to be the closest to the primary photon echo, which is most often used in optical echo spectroscopy.
In this case, the echo signal is emitted in an uninverted atomic medium in a similar way to the CRIB protocol, without the appearance of additional quantum noise.
An important advantage of ROSE protocol in comparison with other QM echo-protocols is the possibility of using a long natural lifetime of optical coherence and a higher optical density of the resonant transition.
These properties of the ROSE protocol make it promising for further development.

The ROSE protocol was experimentally implemented in a bulk crystals \cite{Damon2011,Gerasimov2017OS-ROSE,Liu2024}, in a waveguide \cite{Liu2020LWG,Moiseev2023} and in an impedance-matched cavity \cite{Minnegaliev2018,Minnegaliev2021}.
The retrieval efficiency of $\geq$  40 $\%$  was achieved in the works \cite{Dajczgewand2014, Minnegaliev2022}.
However, a significant problem  of the ROSE protocol is the difficulty of implementing $\pi$ pulses in atomic media, which causes noise in the emitted photon echo signal \cite{Minnegaliev2021, Minnegaliev2022,Minnegaliev2023, Bonarota2014}.

\begin{figure}
%\begin{center}\vspace{1cm}
%\includegraphics[width=1\linewidth]{Fig12-ROSE-protocol.png}
%\includegraphics[width=1\linewidth]{Fig12-ROSE-protocol.eps}
\includegraphics[width=1\linewidth]{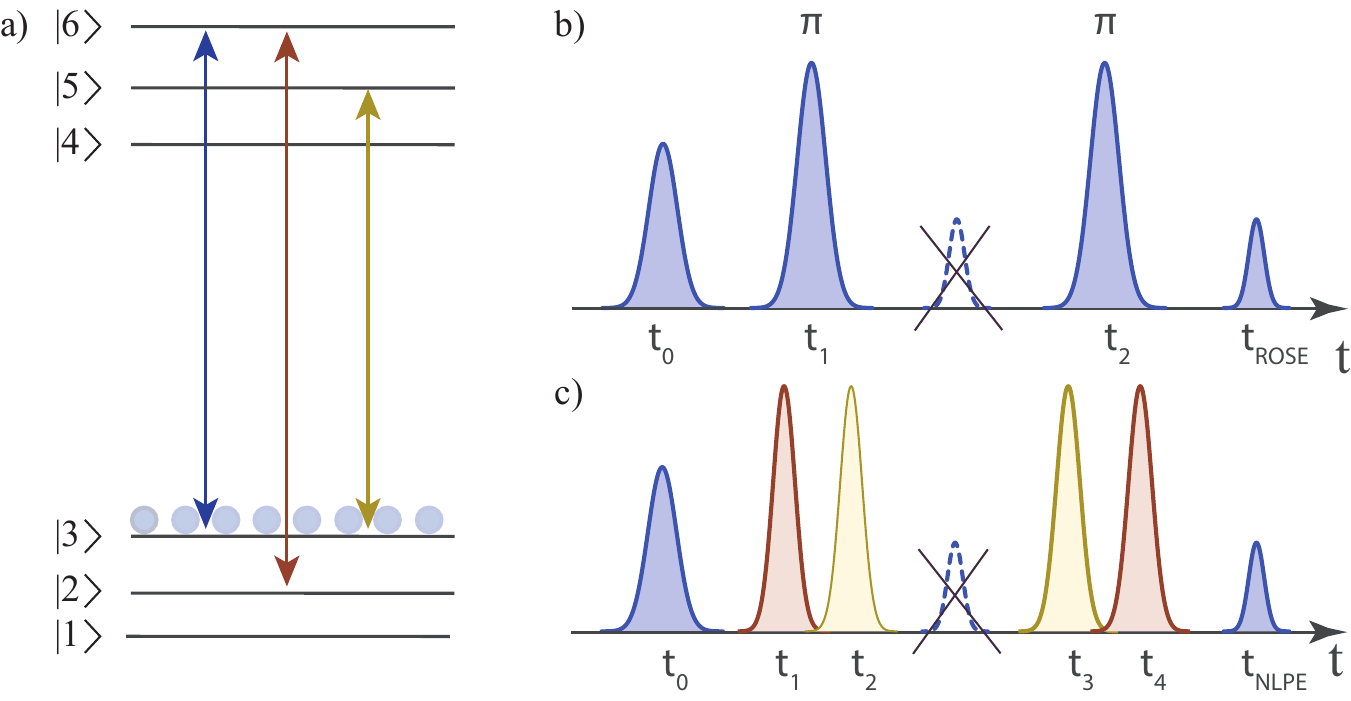}
\caption{ a) Schematic energy level structure and optical transitions of rare-earth ion in crystal.
Pulse sequences in two-level (b) and four-level (c) ROSE (NLPE) protocols.}
\label{ROSE-seq}
\end{figure}

Recent work \cite{Ma2021NLPE} demonstrated the possibility of significantly suppressing the optical noise in ROSE protocol through the use of a 4-level atomic scheme. 
In this protocol, called by noiseless photon-echo (NLPE) protocol, laser $\pi$ pulses acted on an adjacent optical transition, and the resulting optical noise was suppressed using a filter crystal (see Fig.\ref{ROSE-seq}). 
Thus, it was possible to increase the SNR to 42.5 with a storage time of 22.5 $\mu$s and efficiency  $\eta\approx 6.4$\%. 
Despite a significant increase in SNR, in order to achieve a practically significant value of QM efficiency, it is necessary to improve the implementation of rephasing control laser pulses, so that their pulse area will be close to $\pi$.

At the same time, it should be noted that the ROSE protocol differs significantly from the CRIB, GEM and AFC protocols by using intense control laser pulses at resonant atomic transitions of high optical depth.
Presence of high optical depth leads to nonlinear propagation effects of control laser pulses, leading to considerable  changes of its parameters and quantum noises in ROSE signal.
Studying these effects and emerging problems is very important to find optimal conditions and ways to implement ROSE protocols.

The nonlinear effects of coherent control and propagation of control pulses are enhanced when the pulse areas of these pulses become different from $\pi$. 
The consequence of these effects may be a significant deterioration of the properties of the ROSE protocol, even when working with  weak signal pulses.
It should be noted that this situation remains typical for the current level of experimental implementation of optical QM due to the difficulties of implementing laser $\pi$ pulses.

By using photon echo area theorem \eqref{theta}, we focus below on the study of nonlinear propagation effects of controlling laser pulses in  experimentally implemented versions of the ROSE protocol \cite{Minnegaliev2022,Minnegaliev2023, Damon2011} using a two-level system of atoms (see Fig.\ref{ROSE-seq}).
It is also worth noting that the approach based on the use of the area theorem allows us to obtain basic information about the properties of photon echo signals when the spectral width of the signal pulses is much smaller than the inhomogeneous broadening of the atomic transition.
It has been demonstrated in recent works 
for ROSE protocol in a two-level medium with a forward geometry of the echo signal emission 
\cite{Minnegaliev2022,Minnegaliev2023}, in a  cavity \cite{Minnegaliev2021} and in a single mode waveguide \cite{Moiseev2023}.

Here we are particularly interested in the nonlinear properties of the two-level ROSE protocol, which are manifested by the use of control laser pulses, the pulse areas of which begin to differ from $\pi$.
The studied ROSE protocols \cite{Minnegaliev2022,Ma2021NLPE} were  implemented  in the forward geometry of the echo signal emission. 
In this case, the control laser pulses  propagate in the opposite direction to the signal pulse, which makes it possible to noticeably attenuate the optical noise caused by the action of laser pulses.
In experiments \cite{Minnegaliev2022,Ma2021NLPE}, weak coherent signal pulse were used which can be characterize by  small input pulse area $\theta_s(z)=\theta_s(0)e^{-\frac{\alpha}{2}z}$ (where $\theta_s(0)\ll1$).
Therefore, ignoring the population of the excited optical level after the absorption of the signal pulse, we get the solutions for pulse areas of intensive first and second control laser pulses  by using McCall-Hahn area theorem \cite{McCall69,Hahn1971}:

\fla{
&\tan[\frac{\theta_1(z)}{2}]\cong\beta_1e^{-\alpha z/2},\nonumber
\\
&\tan[\frac{\theta_2(z)}{2}]\cong\beta_2\left(\frac{1+\beta_1^2}{e^{\alpha z}+\beta_1^2}\right)e^{\alpha z/2},
\label{theta_controls}
}
where $\beta_{1,2}=\tan[\frac{\theta_{1,2}(0)}{2}]$ and 
as can be seen from the solution \eqref{theta_controls}, if the pulse area of the first pulse is close to $\pi$ and $\beta_1^2\gg e^{\alpha z}$, the second pulse experiences amplification during  its propagation in the medium. 

The solutions \eqref{theta_controls} can be used for studies of the protocol efficiency in the case where the spectrum of laser pulses is much narrower than inhomogeneous broadening of the atomic transition, but wider than the spectrum of the signal pulse \cite{Minnegaliev2021}. 
Moreover, the solutions remain manageable when using chirped laser pulses \cite{Kaup1977}, which allow a significant increase in the working spectral range for the perfect laser control of atomic coherence.
We interested by the case of weak signal pulse pulse and intensive two pulse sequence of control laser  pulses (see Fig.\ref{ROSE-seq}), where we get the following relations for the resonant atomic coherence and inversion  \cite{Minnegaliev2021}:

\fla{
P_e(0,z)=&\Gamma(t_e)e^{-\alpha_R(0)z/2} \theta_s(0)\cdot
\nonumber
\\
&\sin^2{\frac{\theta_1(z)}{2}} \sin^2{\frac{\theta_2(z)}{2}},
}
\fla{
W_e(0,z)=
-\cos{\theta_1(z)} \cos{\theta_2(z)},
}

The formal solution of Eq.\eqref{theta} for the pulse area $\theta_R(z)$ of forward ROSE protocol is:

\fla{
&\theta_R(z)=
2\arctan\big\{\frac{1}{2}\Gamma(t_e)\theta_s(0)\alpha_R(0)\cdot
\nonumber
\\
&\int_{0}^z dz' P_e(0,z')\exp\{\frac{\alpha_R(0)}{2}\int_{z'}^z dz"W_e(0,z") \}\big\},
\label{theta-3}
}
which we use in further  analysis.

To study the influence of deviation of control pulse areas from $\pi$,  in the theoretical description we limit ourselves to the case when the two control laser pulses propagate parallel to the first signal pulse and the photon echo.
Substituting expressions  \eqref{theta_controls} in \eqref{theta-3} after calculating the integrals, we get
\fla{
\theta_e(z)=2\arctan\big\{\frac{1}{2}\Gamma(t_e)\theta_{s}(0)\Phi(z) A(z)\big\},
\label{ROSE-solution}
}
where 
\fla{
A(z)=&
\big\{\alpha_R(0)z+\frac{(1-e^{\alpha_R(0)z})}
{(1+\beta_1^{2})[1+\beta_1^{-2}e^{\alpha_R(0)z}]}
\nonumber
\\
+&\ln\big[\frac{1+\beta_1^{-2}}{(1+\beta_1^{-2}e^{\alpha_R(0)z})}\big]
\big\},
}
\fla{
\Phi(z)=&\frac{e^{\frac{1}{2}\alpha_R(0)z}\beta_2^2(1+\beta_1^2)^2(\beta_1^2+e^{\alpha_R(0)z})}
{\beta_1^2[\beta_1^4+Me^{\alpha_R(0)z}+e^{2\alpha_R(0)z} ]},
}
where $M=\beta_2^2(1+\beta_1^4)+2\beta_1^2(1+\beta_2^2)$ and we assumed that the primary echo signal is suppressed by the additional controlled dephasing \cite{Alqedra2024}.

\begin{figure}
%\begin{center}\vspace{1cm}
%\includegraphics[width=1\linewidth]{F13.eps}
\includegraphics[width=1\linewidth]{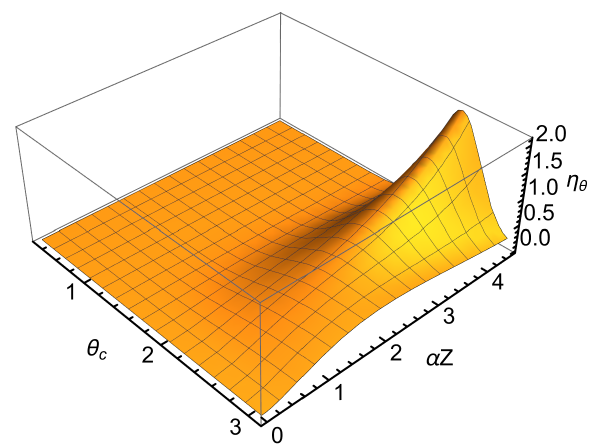}
\caption{ Efficiency in forward two-level ROSE protocol 
$\eta_{\theta}(\theta_e(z))=|\theta_e(z)/\theta_s(0)|^2$
as a function of optical depth $\alpha_R(0)Z$ and pulse areas of control laser pulses $\theta_1(0)=\theta_2(0)=\theta_c$ calculated by using exact solution Eq.\eqref{ROSE-solution}, it is assumed $\theta_e(z)\ll\pi,\theta_s(0)\ll\pi$.}
\label{ROSE-forward eff 2}
\end{figure}

Together with the backward CRIB protocol \eqref{eq-7}, \eqref{CRIB-forw}, the solution \eqref{ROSE-solution} is the third example of an exactly solvable task of photon echo QM protocols under the conditions of strong nonlinear light-atoms interactions. 
The question arises - what will be the behavior of the  ROSE signal intensity with deviation of the pulse area of control laser pulses from $\pi$ in the optically depth medium? 
The  depicted in Fig. \ref{ROSE-forward eff 2} analytical solution \eqref{ROSE-solution} shows that with a significant deviation in the areas of control pulses from $\pi$ can lead to strong amplification of the echo signal in the optical depth medium $\alpha_R(0)z>2.5$.
The maximum growth occurs at $\theta_{c}\sim0.8 \pi$.
However the signal amplification is observed even under conditions of the enlightenment of the atomic medium, which cannot yet act as an amplifier. 
The increase in the echo signal amplitude is a manifestation of the nonlinear coherent interaction of light with atoms in such a medium. 
At the same time, the appearance of optical quantum noise is the result of spontaneous transitions of atoms to the ground state. 
The level of such quantum noise will remain weak with a small deviation of the pulse area of the control laser pulses from $\pi$. 
A detailed study of the level of quantum noise will require special research based on the application of strict quantum theory.
It is also seen that by choosing an optical depth $\alpha_R(0) L =2$, there is a maximum echo signal amplitude with a minimum gain for a noticeable deviation of the pulse areas of the control laser fields from $\pi$, which ensures a low level of optical quantum noise, respectively.

Thus, the solution obtained using the area theorem, makes it possible to determine the required accuracy of choosing the pulse area of control pulses used in an optically dense medium. 
A similar theoretical study can be carried out for a 4-level ROSE scheme, where, however, obtaining an closed analytical solution requires a special study, this also concerns the analysis of the backward geometry of ROSE protocol schemes.

It is also important when implementing the spin-wave  protocol,  in particular, directly using the four-level ROSE protocol.
Currently, different groups use pulses of a special time and frequency dependence, for example complex hyperbolic secant pulse \cite{Silver1985,Roos2004} and hyperbolic-square-hyperbolic pulse \cite{Tian2011} which are especially effective given the pulse duration limitation.
In experimental works \cite{Damon2011,Bonarota2014, Ma2021NLPE,Businger2020, Minnegaliev2021,Minnegaliev2022}, a pulse area in the range from 0.7$\pi$ to 0.9$\pi$ was achieved.
In this case the area theorem approach can be useful as well.

\section{Multimode repeater with absorptive memory}

In quantum communication channels, the error probability scales exponentially with the channel length.
To overcome this limitation a quantum repeater scheme was proposed in the work \cite{1998-PRL-BDCZ}. 
Later it was theoretically shown \cite{2007-PRL-Simon,Sangouard2008} that the use of a multimode optical quantum memory can increase the rate of generation of quantum entanglement by several orders of magnitude between quantum repeater nodes \cite{Sangouard2011,2012-RMP-Multiphoton} and reduce the requirements for the entanglement storage time. 
Instead of generating entanglement between an atom and light, as in the DLCZ protocol \cite{2001-Nature-DLCZ}, it was proposed to use an external source of photon pairs (biphotons) in an entangled quantum state. 
The requirement for the QM in this version of the repeater is the ability to record, store and retrieve several independent modes, for example, temporal, spatial or frequency modes, to ensure the appropriate multiplexing. 

Another advantage of the multimode quantum repeater is the use of two-photon interference in the Bell state measurement nodes, in contrast to the single-photon interference used in the DLCZ protocol. 
Single-photon interference imposes additional requirements for strict phase stability of the control lasers and the length of the fiber (interferometer), which is extremely difficult in practical applications using long distances to transmit quantum information. 
An external source of biphotons is often a nonlinear crystal in which the spontaneous parametric down conversion (SPDC) effect is realized. 
This effect is a nonlinear process of interaction of radiation with an anisotropic crystal, in which a pair of photons is spontaneously born from one photon of intense pump radiation \cite{2021-RevSciInstr-Anwar}. 
In this case, quantum entanglement can be encoded into various degrees of freedom of biphotons, such as polarization, frequency, spatial or temporal modes, orbital angular momentum, as well as into temporal bins (wave packets of single-photon states) \cite{2021-RevSciInstr-Anwar}.

So far the main optical QM scheme used for quantum teleportation of the quantum state of a photon and a qubit in crystals doped with REIs is the AFC protocol.
In the work \cite{Bussieres2014} teleportation of the polarization state of a telecom-wavelength ($\lambda \sim$ 1.43 $\mu$m) photon onto the state of a polarization preserving QM cell.
Entanglement is established between a Nd$^{3+}$:Y$_2$SiO$_5$ crystal ($\lambda \sim$ 883 nm, $\eta _{QM} =$ 5\%, storage time 50 ns), which stores a single photon whose polarization is entangled with a flying photon of telecommunication wavelength. 
%which stores one photon whose polarization is entangled with a flying photon of telecommunication wavelength.
The latter is jointly measured with another flying polarization qubit in the weak coherent state to be teleported, which heralds the teleportation. 
The fidelity of the qubit retrieved from the memory is shown to be $F  >$ 80\%.

Quantum entanglement between two solid-state QM cells was experimentally demonstrated in \cite{2020-PRR-Tittel}. 
Entangled pairs of photons in time-bin modes were used, with photon wavelengths in each pair of 794 nm and 1535 nm. 
They were stored in a Tm$^{3+}$:LiNbO$_{3}$  crystal ($\lambda$ = 794 nm) and in an optical fiber doped with erbium ions Er$^{3+}$ ($\lambda$ = 1535 nm). Quantum entanglement between two different QM cells was obtained with a fidelity of $\sim$93\%. 
Due to the low efficiency of such QM cells, which was 0.1\% in the erbium-doped fiber with storage time (T$_{st}$) of 6 ns and 0.4\% for the Tm$^{3+}$:LiNbO$_{3}$ crystal (T$_{st}$=32 ns), the probability of entanglement of two QM cells was about 4$\cdot$10$^{-6}$, which is not sufficient for practical use and stimulates further improvement of this QM.

In \cite{2023-NatComm-HDR} quantum teleportation over a distance of 1 km was experimentally realized between an enegy-time entangled photonic qubits at a telecommunication wavelength ($\lambda \sim$ 1.43 $\mu$m) and a photonic qubit with a wavelength of 606 nm, which was stored in a Pr$^{3+}$:Y$_2$SiO$_5$ crystal.
In the case of successful interference of two photons at a telecommunication wavelength with subsequent measurement of the Bell state, the probability of teleportation of the qubit after the QM cell for a short distance of 5 m was 1.2$\cdot$10$^{-2}$, while for a distance of 1 km it decreased slightly and was 7.5$\cdot$10$^{-3}$. 
The frequency of successful measurement of the Bell state was 1 Hz at $\eta _{QM}$ $\approx$ 18\%. 
In this experiment, a feedback system was additionally used, implementing a conditional phase shift of the qubit extracted from the QM, as required by the quantum teleportation protocol and time multiplexing was also demonstrated.
Note that it is possible to improve the parameters of implemented QM and the whole system, such as on-demand  readout of photonic qubit \cite{Rakonjac2021} , fiber integration\cite{Rakonjac2022} , heralded entanglement between two spatially separated QMs \cite{Lago-Rivera2021} and increase in efficiency of QM cell up to 60\% \cite{2023-OExp-Duranti} by application of impedance-matched cavity.

Heralded distribution of quantum entanglement between two QM cells was experimentally implemented in the work \cite{2021-NAture-Liu}. 
The experimental setup consisted of two separate sources of polarization-entangled photon pairs that interfered at a beam splitter and were then projected onto Bell states, resulting in polarization entanglement of the two states that were stored in separate QM cells at a distance of 3.5 m. 
The QM cell was a Nd$^{3+}$:YVO$_4$ crystal ($\lambda$ = 880 nm).
The AFC protocol were implemented in these crystals with a storage time of 56 ns, a bandwidth of 1 GHz, a signal photon recovery efficiency of $\eta _{QM}$ $\approx$ 13\%, and a quantum state recovery fidelity $F \approx$ 96\%. 
Bell states were detected at a frequency of 100 Hz, which predicted entanglement of the two QM cells with each other. 
The final entanglement distribution frequency, taking into account all losses, was 1.1 per hour with a quantum state recovery fidelity $F$ = 80.4\%. 
So far this experiment is the closest to a scalable quantum repeater node. 

The above works used only one degree of freedom (DOF) of photons for entanglement creation. 
They can also be entangled in more than one of their DOF, i.e. hyperentangled, and can share more bits of entanglement \cite{Huang2022, Zeng2024}.
In the work \cite{2015-OPtica-Tiranov} the storage of energy-time and polarization hyperentanglement.  One of the photons were stored in a Nd$^{3+}$:Y$_2$SiO$_5$ crystal ($\lambda \sim$ 883 nm), while the other photon has a telecommunication wavelength suitable for transmission in optical fiber.

It should be noted that recently the components of quantum repeater nodes have also been actively developed with importance for practical application.
In the work \cite{2023-NatComm-Jiang} storage  in Er$^{3+}$:Y$_2$SiO$_5$ crystal of the entangled state of two telecom photons generated from an integrated photonic chip was implemented.
The erbium doped fiber was used as a quantum memory cell in the works \cite{Wei2024,Saglamyurek2016} to store photons at telecommunications wavelengths, making it easier to integrate into fiber communication lines.
Another way is to connect fibers to waveguides in crystals in which QM is implemented as was demonstrated in \cite{Zhang2023, Rakonjac2022,Liu2022}.

Simultaneous achievement of the required values  of the QM characteristics (telecom operating wavelength, high storage bandwidth and storage efficiency, long storage time and noiseless) has not yet been implemented in any of the protocols. 
Below we summarize the best achieved parameters in the experimental implementation of photon echo based optical quantum memory.
In the ROSE QM a relatively high efficiency ($>$40\%) at telecom wavelength can be achieved with a storage time of tens of $\mu$s \cite{Dajczgewand2014,Minnegaliev2022}. 
The ROSE protocol compares favorably with other QM protocols in its relative simplicity, but the complexity of creating effective rephasing pulses as close as possible to $\pi$ leads to a decrease in efficiency and the appearance of optical quantum noise greatly exceeding the single-photon level.
For example, in a modified ROSE protocol with four control pulses and the use of two additional  levels \cite{Ma2021NLPE}, it was possible to increase the signal-to-noise ratio to $\sim$42 with a storage time of 22.5 $\mu$s of a signal pulse containing an average of 1.17 photons. 
However, the use of additional $\pi$ pulses led to a decrease in efficiency to 6.4$\%$.

CRIB, GEM and AFC protocols of QM are low-noise, but their implementation is complicated by the necessity of preparation of a narrow spectral component or a comb of such components inside the inhomogeneously broadened line of the optical transition. 
In \cite{Hedges2010}  an efficiency of $\sim$69$\%$ was achieved on a Pr$^{3+}$:Y$_{2}$SiO$_{5}$ crystal for storing quantum states of light with a low noise level. 
The implementation of the GEM protocol by using the Zeeman effect in a switchable magnetic field gradient was demonstrated in works in three-level atomic gases on a non-resonant Raman transition \cite{Hosseini2011, Hosseini2009, Cho2016, Sparkes2012}. 
In these works, rubidium vapor gas was used and an efficiency of 87$\%$ was achieved, which remains the best efficiency in the implementation of optical QM. 
At the same time, the spectral width of such memory was only about 100 kHz.

AFC protocol allows a significant increase in the working spectral width and, accordingly, an increase in the information capacity of the QM.
The AFC protocol was experimentally implemented for different crystals with REIs with different optical transition wavelengths \cite{Holzapfel2020, Stuart2021, Cruzeiro2018, Akhmedzhanov2023, Rielander2014 }. 
 In AFC-protocol an efficiency of 62\% was achieved \cite{Duranti2023}, a storage time ($T_{st}$) of 1 hour \cite{Ma2021}, fidelity of  $>$ 99\% \cite{Zhu2022,Zhou2012} (for efficiency of 11 and 7$\%$), the storage of more than 1000 temporal light modes \cite{Businger2022,Wei2024,Bonarota2011} and a large operating spectral width ($\Delta_{0} > 6$ GHz)  were demonstrated \cite{Saglamyurek2011,Wei2024, 2020-PRR-Tittel} but with efficiency of $<10\%$.

We also draw the reader’s attention to recent reviews devoted to the issues on theoretical and experimental implementation of quantum repeater nodes in other physical systems \cite{Wei2022}, as well as an analysis of requirements of qunatum memory efficiency and storage time \cite{2022-AVSquantum-Semenenko}. 

\section{Conclusion}

We have considered the basic photon echo QM protocols, analyzed their main properties and key problems in achieving high efficiency for  broadband signal fields.
Using the new obtained analytical results we analyzed effects of spectral dispersion on the efficiency of QM echo-protocols and  
nonlinear effects of the atomic interactions with intensive signal and control light pulses. 
The presented analytical results  and performed analysis  made it possible to find requirements for the atomic and light parameters in the efficient storage of signal fields.
Basic properties and current problems of QM echo-protocols were discussed together with recent experimental results  demonstrated significant progress in improving the performance of the studied echo-protocols. 
Concluding the analysis of basic echo protocols of QM, it is worth noting the need for further development of the approaches to their implementation. 
In this regard, it is of interest to use pre-created long-lived macroscopic coherence in the system of atoms \cite{Moiseev2024_OQM_MC}, in which the possibility of suppressing quantum noise and dynamic programming of storage time appears.

Description of optical quantum memory on atomic ensembles requires taking into account a large number of parameters, including both the parameters of the macroscopic system and the microscopic parameters of atoms, without which it is difficult to expect to achieve high efficiency of the implemented protocols.
At present, the fidelity in the implementation of quantum memory protocols remains not as high as  necessary for the implementation of quantum computing, which is due to the complexity of describing all the factors involved in the dynamics of the functioning of quantum memory. 
The description of existing protocols is still carried out using model approaches to describing the dynamics of the behavior of atoms and light, which is in satisfactory agreement with existing experimental data within the framework of experimental accuracy. 
It should be expected that subsequent progress in improving experimental studies of optical quantum memory will lead to the need for a more detailed and more accurate description of the quantum dynamics of light and atoms interacting with it. 
These issues currently arise when it comes to the need for a more detailed description of the behavior of atoms experiencing strong interactions with neighboring atoms and the environment. 
To describe the quantum dynamics of such atomic systems, the use of a quantum computer will soon become an urgent task. 
It is also necessary to expect the receipt of experimental data that claim to be of higher accuracy, which will make it important to build more complex theories and the need for powerful computing resources for their use and verification.

Among the most important tasks of the near future, which it is important to solve on the way to improving the basic parameters of broadband QM on photon echo, we would like to draw the attention of researchers to the solution of the following tasks.

1. Ensuring a high degree of temporal reversibility of protocols and compensation for the influence of spectral dispersion to achieve high efficiency and fidelity of signal pulse retrieval.

2. Development of experimental methods for the realization of laser $\pi$ pulses in the excitation of optically dense atomic ensembles.

3. Improvement of methods for the preparation of the initial state (high-precision depopulation of spin sublevels) of rare earth ions.

4. Further development of  dynamical-decoupling rf- and microwave pulse sequences providing high preserving of coherency and techniques of depopulation of spin sub-levels for high enough SNR. Now the best SNR is 7.4 for 20 ms storage time using AFC-protocol \cite{Ortu2022}.

At the same time, further improvement of the QM echo protocols should be expected using optical resonators \cite{Moiseev2010cavity,Afzelius2010}.
Cavity assisted QMs will significantly reduce the requirements for the optical depth of the resonant transition, involve off-resonant Raman interaction for direct mapping to long-lived states of REI \cite{Moiseev2013PRA,Kalachev2013}, as well as its use in various integrated circuits \cite{Meng2024,Zhou2023,Labonte2024}.
Along this path, there are still many tasks to be solved in the development of new QM resonator  schemes \cite{EMoiseev2021,Perminov2018} for controlling the interaction of photons with coherent atomic ensembles.
The development of broadband QM using multi-resonator systems is also of a large interest here \cite{Moiseev_2017, Matanin2023,Perminov2023}.

When implementing optical quantum memory protocols in concentrated atomic ensembles, 
significant development of the theory is already important problem, which takes into account the influence of interatomic interactions, the resulting features of atomic relaxation, the manifestation of the local Lorentz field in the dynamics of atoms, the propagation of light fields in waveguide structures, etc.
These questions arise in specific experiments, the detailed analysis of which will require further development of the theory, the application of which will require the development of numerical methods for analyzing the studied quantum memory protocols.
We also draw the reader's attention to recent reviews devoted to the issues of experimental implementation of optical QM.
\cite{Guo2023,Lei2023,Pettit2023}
and to the reviews on first  experimental results obtained in integrated QM using  REI doped crystals 
 \cite{Zhou2023,Labonte2024},
 where the influence of spectral dispersion and nonlinear interaction of light pulses with atomic systems also require special study\cite{EMoiseev2021}.

\subsection*{Acknowledgments}
The authors greatly thank Dr. N. M. Arslanov for useful discussions. This work was carried out with the support of the Ministry of Science and Higher Education of the Russian Federation (Register No. 121020400113-1).

 \section*{References}

\bibliographystyle{iopart-num} 
\bibliography{ref}

\providecommand{\newblock}{}
\begin{thebibliography}{100}
\expandafter\ifx\csname url\endcsname\relax
  \def\url#1{{\tt #1}}\fi
\expandafter\ifx\csname urlprefix\endcsname\relax\def\urlprefix{URL }\fi
\providecommand{\eprint}[2][]{\url{#2}}

\bibitem{Sangouard2011}
Sangouard N, Simon C, de~Riedmatten H and Gisin N 2011 {\em Rev. Mod. Phys.\/} {\bf 83}(1) 33--80

\bibitem{Kok2007}
Kok P, Munro W~J, Nemoto K, Ralph T~C, Dowling J~P and Milburn G~J 2007 {\em Reviews of Modern Physics\/} {\bf 79} 135--174 ISSN 0034-6861

\bibitem{Stevenson2014}
Stevenson R~N, Hush M~R, Carvalho A~R~R, Beavan S~E, Sellars M~J and Hope J~J 2014 {\em New Journal of Physics\/} {\bf 16} 033042

\bibitem{Manukhova2017}
Manukhova A~D, Tikhonov K~S, Golubeva T~Y and Golubev Y~M 2017 {\em Phys. Rev. A\/} {\bf 96}(2) 023851

\bibitem{Chen2023}
Chen D~L, Zhou Z~Q, Li C~F and Guo G~C 2023 {\em Phys. Rev. A\/} {\bf 107}(4) 042619

\bibitem{Clauser1969}
Clauser J~F, Horne M~A, Shimony A and Holt R~A 1969 {\em Phys. Rev. Lett.\/} {\bf 23}(15) 880--884

\bibitem{Brunner2014}
Brunner N, Cavalcanti D, Pironio S, Scarani V and Wehner S 2014 {\em Rev. Mod. Phys.\/} {\bf 86}(2) 419--478

\bibitem{Mol2023}
Mol J~M, Esguerra L, Meister M, Bruschi D~E, Schell A~W, Wolters J and Worner L 2023 {\em Quantum Science and Technology\/} {\bf 8} 024006

\bibitem{Hosseini2012}
Hosseini M, Sparkes B~M, Campbell G~T, Lam P~K and Buchler B~C 2012 {\em Journal of Physics B: Atomic, Molecular and Optical Physics\/} {\bf 45} 124004 ISSN 0953-4075

\bibitem{Moiseev2016}
Moiseev E~S and Moiseev S~A 2016 {\em Journal of Modern Optics\/} {\bf 63} 2081--2092 ISSN 0950-0340

\bibitem{Chen2021}
Chen K~C, Dai W, Errando-Herranz C, Lloyd S and Englund D 2021 {\em PRX Quantum\/} {\bf 2} 030319 ISSN 2691-3399

\bibitem{Lvovsky2009}
Lvovsky A~I, Sanders B~C and Tittel W 2009 {\em Nature Photonics\/} {\bf 3} 706--714 ISSN 17494885 (\textit{Preprint} \eprint{arXiv:1002.4659v3})

\bibitem{Tittel2009}
Tittel W, Afzelius M, Chaneli{\'{e}}re T, Cone R, Kr{\"{o}}ll S, Moiseev S and Sellars M 2009 {\em Laser and Photonics Reviews\/} {\bf 4} 244--267 ISSN 18638880

\bibitem{Bussieres2013}
Bussi{\`{e}}res F, Sangouard N, Afzelius M, Riedmatten H~D and Tittel W 2013 {\em Journal of Modern Optics\/} {\bf 60} 1519--1537

\bibitem{Heshami2016a}
Heshami K, England D~G, Humphreys P~C, Bustard P~J, Acosta V~M, Nunn J and Sussman B~J 2016 {\em Journal of Modern Optics\/} {\bf 63} 2005--2028 ISSN 0950-0340 (\textit{Preprint} \eprint{1511.04018})

\bibitem{Chaneliere2018}
Chaneli{\`{e}}re T, H{\'{e}}tet G and Sangouard N 2018 Chapter two - quantum optical memory protocols in atomic ensembles {\em Advances In Atomic, Molecular, and Optical Physics\/} vol~67 ed Arimondo E, DiMauro L~F and Yelin S~F (Academic Press) pp 77--150

\bibitem{Hua2018}
Hua Y~L, Zhou Z~Q, Li C~F and Guo G~C 2018 {\em Chinese Physics B\/} {\bf 27} ISSN 20583834

\bibitem{Guo2023}
Guo M, Liu S, Sun W, Ren M, Wang F and Zhong M 2023 {\em Frontiers of Physics\/} {\bf 18} 1--19 ISSN 20950470

\bibitem{Zhou2023}
Zhou Z~Q, Liu C, Li C~F, Guo G~C, Oblak D, Lei M, Faraon A, Mazzera M and de~Riedmatten H 2023 {\em Laser and Photonics Reviews\/} {\bf 17} 1--28 ISSN 18638899

\bibitem{Lei2023}
Lei Y, Asadi F~K, Zhong T, Kuzmich A, Simon C and Hosseini M 2023 {\em Optica\/} {\bf 10} 1511--1528

\bibitem{Hahn1950}
Hahn E~L 1950 {\em Phys. Rev.\/} {\bf 80}(4) 580--594

\bibitem{Kopvillem1963}
Kopvillem U and Nagibarov V 1963 {\em Fiz. Metal. i Metalloved.\/} {\bf 15} 313

\bibitem{Kurnit1964}
Kurnit N~A, Abella I~D and Hartmann S~R 1964 {\em Phys. Rev. Lett.\/} {\bf 13}(19) 567--568

\bibitem{GORIN2006}
Gorin T, Prosen T~Seligman T~H and Znidaric M 2006 {\em Physics Reports\/} {\bf 435} 33--156 ISSN 0370-1573

\bibitem{Wisniacki2012}
Wisniacki A 2012 {\em Scholarpedia\/} {\bf 7} 11687 ISSN 1941-6016

\bibitem{Moiseev2001}
Moiseev S~A and Kr{\"{o}}ll S 2001 {\em Physical Review Letters\/} {\bf 87} 173601 ISSN 10797114

\bibitem{Moiseev2004}
Moiseev S~A and Noskov M~I 2004 {\em Laser Physics Letters\/} {\bf 1} 303--310 ISSN 16122011

\bibitem{Moiseev2003}
Moiseev S~A, Tarasov V~F and Ham B~S 2003 {\em Journal of Optics B: Quantum and Semiclassical Optics\/} {\bf 5} S497--S502 ISSN 1464-4266

\bibitem{Moiseev2004a}
Moiseev S~A and Ham B~S 2004 {\em Physical Review A\/} {\bf 70} 1--16 ISSN 10502947

\bibitem{Kraus2006}
Kraus B, Tittel W, Gisin N, Nilsson M, Kr{\"{o}}ll S and Cirac J~I 2006 {\em Physical Review A\/} {\bf 73} 020302 ISSN 10502947 (\textit{Preprint} \eprint{0502184})

\bibitem{Alexander2006}
Alexander A~L, Longdell J~J, Sellars M~J and Manson N~B 2006 {\em Physical Review Letters\/} {\bf 96} 043602 ISSN 0031-9007

\bibitem{Lauritzen2010}
Lauritzen B, Min{\'{a}}ř J, de~Riedmatten H, Afzelius M, Sangouard N, Simon C and Gisin N 2010 {\em Physical Review Letters\/} {\bf 104} 080502 ISSN 0031-9007

\bibitem{Sangouard2007}
Sangouard N, Simon C, Afzelius M and Gisin N 2007 {\em Physical Review A\/} {\bf 75} 032327 ISSN 1050-2947

\bibitem{Gorshkov2007}
Gorshkov A~V, Andr{\'{e}} A, Lukin M~D and S{\o}rensen A~S 2007 {\em Physical Review A\/} {\bf 76} 033804 ISSN 1050-2947

\bibitem{Moiseev2011}
Moiseev S~A and Tittel W 2011 {\em New Journal of Physics\/} {\bf 13} 063035 ISSN 1367-2630

\bibitem{Moiseev2013}
Moiseev E~S and Moiseev S~A 2013 {\em New Journal of Physics\/} {\bf 15} 105005 ISSN 1367-2630

\bibitem{Nilsson2004}
Nilsson M, Rippe L, Kr\"oll S, Klieber R and Suter D 2004 {\em Phys. Rev. B\/} {\bf 70}(21) 214116

\bibitem{Hosseini2011}
Hosseini M, Sparkes B, Campbell G, Lam P and Buchler B 2011 {\em Nature Communications\/} {\bf 2} 174 ISSN 2041-1723

\bibitem{Moiseev2023}
Moiseev S~A, Minnegaliev M~M, Moiseev E~S, Gerasimov K~I, Pavlov A~V, Rupasov T~A, Skryabin N~N, Kalinkin A~A and Kulik S~P 2023 {\em Physical Review A\/} {\bf 107} 043708 ISSN 2469-9926

\bibitem{Merkel2021}
Merkel B, {Cova Fari{\~{n}}a} P and Reiserer A 2021 {\em Physical Review Letters\/} {\bf 127} 030501 ISSN 0031-9007

\bibitem{Moiseev2015}
Moiseev S~A and Skrebnev V~A 2015 {\em Journal of Physics B: Atomic, Molecular and Optical Physics\/} {\bf 48} ISSN 13616455

\bibitem{Moiseev2015PRA}
Moiseev S~A and Skrebnev V~A 2015 {\em Phys. Rev. A\/} {\bf 91}(2) 022329

\bibitem{Waeber2019}
Waeber A~M, Gillard G, Ragunathan G, Hopkinson M, Spencer P, Ritchie D~A, Skolnick M~S and Chekhovich E~A 2019 {\em Nature Communications\/} {\bf 10} 3157 ISSN 2041-1723

\bibitem{Minnegaliev2019}
Minnegaliev M~M, Urmancheev R~V, Skrebnev V~A and Moiseev S~A 2019 {\em Optics and Spectroscopy\/} {\bf 126} 1--5 ISSN 15626911

\bibitem{Heinze2013}
Heinze G, Hubrich C and Halfmann T 2013 {\em Phys. Rev. Lett.\/} {\bf 111}(3) 033601

\bibitem{Zhong2015}
Zhong M, Hedges M~P, Ahlefeldt R~L, Bartholomew J~G, Beavan S~E, Wittig S~M, Longdell J~J and Sellars M~J 2015 {\em Nature\/} {\bf 517} 177--180 ISSN 0028-0836 (\textit{Preprint} \eprint{1411.6758})

\bibitem{Rancic2017}
Ran{\v{c}}i{\'{c}} M, Hedges M~P, Ahlefeldt R~L and Sellars M~J 2017 {\em Nature Physics\/} {\bf 14} 50--54 ISSN 1745-2473 (\textit{Preprint} \eprint{1611.04315})

\bibitem{Hain2022}
Hain M, Stabel M and Halfmann T 2022 {\em New Journal of Physics\/} {\bf 24} 023012 ISSN 1367-2630

\bibitem{Ma2021}
Ma Y, Ma Y~Z, Zhou Z~Q, Li C~F and Guo G~C 2021 {\em Nature Communications\/} {\bf 12} 2381 ISSN 2041-1723 (\textit{Preprint} \eprint{2012.14605})

\bibitem{Sangouard2007b}
Sangouard N, Simon C, Afzelius M and Gisin N 2007 {\em Phys. Rev. A\/} {\bf 75}(3) 032327

\bibitem{Moiseev2007b}
Moiseev S~A 2007 {\em Journal of Physics B: Atomic, Molecular and Optical Physics\/} {\bf 40} 3877

\bibitem{Ham2018}
Ham B~S 2018 {\em Scientific Reports\/} {\bf 8} 10675 ISSN 2045-2322

\bibitem{Gorshkov2007PRL}
Gorshkov A~V, Andr{\'{e}} A, Fleischhauer M, S{\o}rensen A~S and Lukin M~D 2007 {\em Physical Review Letters\/} {\bf 98} 123601 ISSN 0031-9007

\bibitem{Autler1955}
Autler S~H and Townes C~H 1955 {\em Physical Review\/} {\bf 100} 703--722 ISSN 0031-899X

\bibitem{Saglamyurek2018}
Saglamyurek E, Hrushevskyi T, Rastogi A, Heshami K and LeBlanc L~J 2018 {\em Nature Photonics\/} {\bf 12} 774--782 ISSN 1749-4885

\bibitem{Vernaz-Gris2018}
Vernaz-Gris P, Tranter A~D, Everett J~L, Leung A~C, Paul K~V, Campbell G~T, Lam P~K and Buchler B~C 2018 {\em Optics Express\/} {\bf 26} 12424 ISSN 1094-4087

\bibitem{Dajczgewand2014}
Dajczgewand J, {Le Gou{\"{e}}t} J~L, Louchet-Chauvet A and Chaneli{\`{e}}re T 2014 {\em Optics Letters\/} {\bf 39} 2711 ISSN 0146-9592

\bibitem{Minnegaliev2022}
Minnegaliev M~M, Gerasimov K~I, Sabirov T~N, Urmancheev R~V and Moiseev S~A 2022 {\em JETP Letters\/} {\bf 115} 720--727 ISSN 0021-3640

\bibitem{Moiseev1987}
Moiseev S~A 1987 {\em Optics and Spectroscopy (English translation of Optika i Spektroskopiya)\/} {\bf 62} 180--185

\bibitem{Moiseev2004Izv}
Moiseev S~A 2004 {\em Bulletin of the Russian Academy of Sciences: Physics\/} {\bf 68} 1260

\bibitem{Urmancheev2019}
Urmancheev R, Gerasimov K, Minnegaliev M, Chaneli{\`{e}}re T, Louchet-Chauvet A and Moiseev S 2019 {\em Optics Express\/} {\bf 27} 28983 ISSN 1094-4087

\bibitem{Moiseev2020}
Moiseev S~A, Sabooni M and Urmancheev R~V 2020 {\em Phys. Rev. Res.\/} {\bf 2}(1) 012026

\bibitem{McCall69}
McCall S~L and Hahn E~L 1969 {\em Physical Review\/} {\bf 183} 457--485

\bibitem{Ablowitz1974}
Ablowitz M~J, Kaup D~J and Newell A~C 1974 {\em Journal of Mathematical Physics\/} {\bf 15} 1852--1858 ISSN 0022-2488

\bibitem{Maimistov1990}
Maimistov A, Basharov A, Elyutin S and Sklyarov Y 1990 {\em Physics Reports\/} {\bf 191} 1--108 ISSN 0370-1573

\bibitem{Moiseev_Urmancheev2022}
Moiseev S~A and Urmancheev R~V 2022 {\em Opt. Lett.\/} {\bf 47} 3812--3815

\bibitem{Minnegaliev2021}
Minnegaliev M~M, Gerasimov K~I, Urmancheev R~V, Zheltikov A~M and Moiseev S~A 2021 {\em Physical Review B\/} {\bf 103} 174110 ISSN 2469-9950

\bibitem{Vinh2021}
Vinh N~T, Tsarev D~V and Alodjants A~P 2021 {\em Journal of Russian Laser Research\/} {\bf 42} 523--537 ISSN 1573-8760

\bibitem{Rupasov1982}
Rupasov V~I 1982 {\em Sov. Phys. JETP [Zh. Eksp. Teor. Fiz. 83, 1711-1723 (1982)]\/} {\bf 56} 989

\bibitem{Hedges2010}
Hedges M~P, Longdell J~J, Li Y and Sellars M~J 2010 {\em Nature\/} {\bf 465} 1052--1056 ISSN 0028-0836

\bibitem{Hosseini2009}
Hosseini M, Sparkes B~M, H{\'{e}}tet G, Longdell J~J, Lam P~K and Buchler B~C 2009 {\em Nature\/} {\bf 461} 241--245 ISSN 0028-0836

\bibitem{Cho2016}
Cho Y~W, Campbell G~T, Everett J~L, Bernu J, Higginbottom D~B, Cao M~T, Geng J, Robins N~P, Lam P~K and Buchler B~C 2016 {\em Optica\/} {\bf 3} 100 ISSN 2334-2536

\bibitem{Saidasheva2006}
Saidasheva I, Arslanov N and Moiseev S 2006 Modeling of photon echo with controlled external field gradient: possibility of high efficient quantum memory. {\em Proceedings of the VI-th International Congress “Basic Problems of Optics”,\/} ed prof Kozlov~SA (St. Petersburg State Univ. of Information Technologies, 16-20 October,2006, St.-Petersburg, Russia) p 127.

\bibitem{Moiseev2008}
Moiseev S~A and Arslanov N~M 2008 {\em Physical Review A\/} {\bf 78} 023803 ISSN 1050-2947

\bibitem{Lukin2003}
Lukin M~D 2003 {\em Rev. Mod. Phys.\/} {\bf 75}(2) 457--472

\bibitem{Lukin2000}
Lukin M~D and Imamo\ifmmode~\breve{g}\else \u{g}\fi{}lu A 2000 {\em Phys. Rev. Lett.\/} {\bf 84}(7) 1419--1422

\bibitem{Wang2006}
Wang Z~B, Marzlin K~P and Sanders B~C 2006 {\em Phys. Rev. Lett.\/} {\bf 97}(6) 063901

\bibitem{Marzlin2010}
Marzlin K~P, Wang Z~B, Moiseev S~A and Sanders B~C 2010 {\em J. Opt. Soc. Am. B\/} {\bf 27} A36--A45

\bibitem{Scherer2012}
He B and Scherer A 2012 {\em Phys. Rev. A\/} {\bf 85}(3) 033814

\bibitem{Bienias_2020}
Bienias P and B{\"{u}}chler H~P 2020 {\em Journal of Physics B: Atomic, Molecular and Optical Physics\/} {\bf 53} 054003

\bibitem{Leung_2022}
Leung A~C, Melody K~S~I, Tranter A~D, Paul K~V, Campbell G~T, Lam P~K and Buchler B~C 2022 {\em New Journal of Physics\/} {\bf 24} 093011

\bibitem{Feizpour2015}
Feizpour A, Hallaji M, Dmochowski G and Steinberg A~M 2015 {\em Nature Physics\/} {\bf 11} 905--909 ISSN 1745-2481

\bibitem{Shapiro2006}
Shapiro J~H 2006 {\em Phys. Rev. A\/} {\bf 73}(6) 062305

\bibitem{Banacloche2010}
Gea-Banacloche J 2010 {\em Phys. Rev. A\/} {\bf 81}(4) 043823

\bibitem{Moiseev2010}
Moiseev S~A and Tittel W 2010 {\em Phys. Rev. A\/} {\bf 82}(1) 012309

\bibitem{MoiseevES2020}
Moiseev E~S, Tashchilina A, Moiseev S~A and Lvovsky A~I 2020 {\em New Journal of Physics\/} {\bf 22} 013014 ISSN 1367-2630

\bibitem{Moiseev2013PRA}
Moiseev S~A 2013 {\em Phys. Rev. A\/} {\bf 88}(1) 012304

\bibitem{Kalachev2013}
Kalachev A and Kocharovskaya O 2013 {\em Phys. Rev. A\/} {\bf 88}(3) 033846

\bibitem{2015-Goldner-Handbook}
Goldner P, Ferrier A and Guillot-No{\"{e}}l O 2015 {Rare Earth-Doped Crystals for Quantum Information Processing} {\em Handbook on the Physics and Chemistry of Rare Earths\/} vol~46 (Elsevier B.V.) pp 1--78 ISBN 9780874216561

\bibitem{Popova2019}
Popova M~N, Klimin S~A, Moiseev S~A, Gerasimov K~I, Minnegaliev M~M, Baibekov E~I, Shakurov G~S, Bettinelli M and Chou M~C 2019 {\em Physical Review B\/} {\bf 99} 235151

\bibitem{2023-Gerasimov-OandS}
Gerasimov K~I, Sabirov T~N, Moiseev S~A, Baibekov E~I, Bettinelli M, Chou M, Yen Y~C and Popova M 2023 {\em Optics and Spectroscopy\/} {\bf 131} 607--613

\bibitem{Gerasimov2024}
Gerasimov K, Baibekov E, Minnegaliev M, Shakurov G, Zaripov R, Moiseev S, Lebedev A and Malkin B 2024 {\em Journal of Luminescence\/} {\bf 270} 120564

\bibitem{MoiseevTittel2010}
Moiseev S~A and Tittel W 2010 {\em Phys. Rev. A\/} {\bf 82}(1) 012309

\bibitem{EMoiseev2021}
Moiseev E~S, Tashchilina A, Moiseev S~A and Sanders B~C 2021 {\em New Journal of Physics\/} {\bf 23} 063071

\bibitem{Tame2013}
Tame M~S, McEnery K~R, {\"{O}}zdemir S~K, Lee J, Maier S~M and Kim S 2013 {\em Nature Physics\/} {\bf 9} 329--340

\bibitem{Mois2010}
Moiseev S~A and Moiseev E~S 2010 {\em Multi mode nano scale raman echo quantum memory\/} vol 26 of NATO Science for Peace and Security Series - D: Information and Communication Security, Kowalik, J. and Horodecki, R. and Kilin, S.Y., editors, Quantum Cryptography and Computing: Theory and Implementation (IOS Press BV)

\bibitem{Novotny2012}
Novotny L and Hecht B 2012 {\em Nano-Optics\/} (Cambridge University Press)

\bibitem{Dubetskii1985}
Dubetskii B~Y and Chebotaev V~P 1985 {\em JETP Lett.\/} {\bf 41} 328

\bibitem{Dubetskii1986}
Dubetskii B~Y and Chebotaev V~P 1986 {\em Bull. Acad. Sci. USSR, Phys. Ser.\/} {\bf 50} 70

\bibitem{DeRiedmatten2008}
de~Riedmatten H, Afzelius M, Staudt M~U, Simon C and Gisin N 2008 {\em Nature\/} {\bf 456} 773--777 ISSN 0028-0836

\bibitem{Afzelius2009}
Afzelius M, Simon C, {De Riedmatten} H and Gisin N 2009 {\em Physical Review A\/} {\bf 79} 052329 ISSN 10502947 (\textit{Preprint} \eprint{0805.4164})

\bibitem{Moiseev2012}
Moiseev S~A and {Le Gou{\"{e}}t} J~L 2012 {\em Journal of Physics B: Atomic, Molecular and Optical Physics\/} {\bf 45} 124003

\bibitem{Arslanov2017}
Arslanov N and Moiseev S 2017 {\em Quantum Electronics\/} {\bf 47} 783

\bibitem{Arslanov2019}
Arslanov N~M and Moiseev S~A 2019 {\em Optics and Spectroscopy\/} {\bf 126} 29--33 ISSN 1562-6911

\bibitem{2024-Arslanov-inprogress}
Arslanov N and Moiseev S 2024 {\em (in progress).\/}

\bibitem{Holzapfel2020}
Holz{\"{a}}pfel A, Etesse J, Kaczmarek K~T, Tiranov A, Gisin N and Afzelius M 2020 {\em New Journal of Physics\/} {\bf 22} 0--13 ISSN 13672630 (\textit{Preprint} \eprint{1910.08009})

\bibitem{Stuart2021}
Stuart J~S, Hedges M, Ahlefeldt R and Sellars M 2021 {\em Physical Review Research\/} {\bf 3} L032054 ISSN 2643-1564 (\textit{Preprint} \eprint{2103.04581})

\bibitem{Cruzeiro2018}
Cruzeiro E~Z, Tiranov A, Lavoie J, Ferrier A, Goldner P, Gisin N and Afzelius M 2018 {\em New Journal of Physics\/} {\bf 20} 053013 ISSN 1367-2630 (\textit{Preprint} \eprint{1712.02682})

\bibitem{Akhmedzhanov2016}
Akhmedzhanov R~A, Gushchin L~A, Kalachev A~A, Korableva S~L, Sobgayda D~A and Zelensky I~V 2016 {\em Laser Physics Letters\/} {\bf 13} 015202 ISSN 1612-2011

\bibitem{Rielander2014}
Riel{\"{a}}nder D, Kutluer K, Ledingham P~M, G{\"{u}}ndogan M, Fekete J, Mazzera M and {De Riedmatten} H 2014 {\em Physical Review Letters\/} {\bf 112} 1--5 ISSN 00319007 (\textit{Preprint} \eprint{1310.8261})

\bibitem{Akhmedzhanov2023}
Akhmedzhanov R~A, Gushchin L~A, Kalachev A~A, Nizov N~A, Nizov V~A, Sobgayda D~A and Zelensky I~V 2023 {\em Laser Physics Letters\/} {\bf 20} 015204 ISSN 1612-2011

\bibitem{Duranti2023}
Duranti S, Wengerowsky S, Feldmann L, Seri A, Casabone B and de~Riedmatten H 2023 {\em arxiv\/}  1--7 (\textit{Preprint} \eprint{2307.03509})

\bibitem{Zhu2022}
Zhu T~X, Liu C, Jin M, Su M~X, Liu Y~P, Li W~J, Ye Y, Zhou Z~Q, Li C~F and Guo G~C 2022 {\em Phys. Rev. Lett.\/} {\bf 128}(18) 180501

\bibitem{Zhou2012}
Zhou Z~Q, Lin W~B, Yang M, Li C~F and Guo G~C 2012 {\em Phys. Rev. Lett.\/} {\bf 108}(19) 190505

\bibitem{Businger2022}
Businger M, Nicolas L, Mejia T~S, Ferrier A, Goldner P and Afzelius M 2022 {\em Nature Communications\/} {\bf 13} 1--8 ISSN 20411723

\bibitem{Wei2024}
Wei S~H, Jing B, Zhang X~Y, Liao J~Y, Li H, You L~X, Wang Z, Wang Y, Deng G~W, Song H~Z, Oblak D, Guo G~C and Zhou Q 2024 {\em npj Quantum Information\/} {\bf 10} 1--8 ISSN 20566387 (\textit{Preprint} \eprint{2209.00802})

\bibitem{Bonarota2011}
Bonarota M, , {Le Gou{\"{e}}t} J~L and Chaneli{\`{e}}re T 2011 {\em New Journal of Physics\/} {\bf 13} 013013

\bibitem{Saglamyurek2011}
Saglamyurek E, Sinclair N, Jin J, Slater J~A, Oblak D, Bussi{\`{e}}res F, George M, Ricken R, Sohler W and Tittel W 2011 {\em Nature\/} {\bf 469} 512--515 ISSN 0028-0836

\bibitem{2020-PRR-Tittel}
Puigibert M~l~G, Askarani M~F, Davidson J~H, Verma V~B, Shaw M~D, Nam S~W, Lutz T, Amaral G~C, Oblak D and Tittel W 2020 {\em Phys. Rev. Res.\/} {\bf 2}(1) 013039

\bibitem{Horvath2021}
Horvath S~P, Alqedra M~K, Kinos A, Walther A, Dahlstr{\"{o}}m J~M, Kr{\"{o}}ll S and Rippe L 2021 {\em Physical Review Research\/} {\bf 3} 1--9 (\textit{Preprint} \eprint{2006.00943})

\bibitem{Craiciu2021}
Craiciu I, Lei M, Rochman J, Bartholomew J~G and Faraon A 2021 {\em Optica\/} {\bf 8} 114 ISSN 23342536 (\textit{Preprint} \eprint{2008.10795})

\bibitem{Corrielli2016}
Corrielli G, Seri A, Mazzera M, Osellame R and de~Riedmatten H 2016 {\em Physical Review Applied\/} {\bf 5} 054013 ISSN 2331-7019 (\textit{Preprint} \eprint{1512.09288})

\bibitem{Rakonjac2021}
Rakonjac J~V, Lago-Rivera D, Seri A, Mazzera M, Grandi S and {De Riedmatten} H 2021 {\em Physical Review Letters\/} {\bf 127} 210502 ISSN 10797114 (\textit{Preprint} \eprint{2106.05079})

\bibitem{Ortu2022}
Ortu A, Holz{\"{a}}pfel A, Etesse J and Afzelius M 2022 {\em npj Quantum Information\/} {\bf 8} 1--7 ISSN 20566387 (\textit{Preprint} \eprint{2109.06669})

\bibitem{Alqedra2024}
Alqedra M~K, Horvath S~P, Kinos A, Walther A, Kr{\"{o}}ll S and Rippe L 2024 {\em Physical Review A\/} {\bf 109} 12607 ISSN 2469-9926 (\textit{Preprint} \eprint{2211.17206})

\bibitem{Businger2020}
Businger M, Tiranov A, Kaczmarek K~T, Welinski S, Zhang Z, Ferrier A, Goldner P and Afzelius M 2020 {\em Physical Review Letters\/} {\bf 124} 53606 ISSN 10797114 (\textit{Preprint} \eprint{1907.11571})

\bibitem{Sabooni2013}
Sabooni M, Li Q, Kr{\"{o}}ll S and Rippe L 2013 {\em Physical Review Letters\/} {\bf 110} 133604 ISSN 00319007 (\textit{Preprint} \eprint{1301.0636})

\bibitem{Jobez2014}
Jobez P, Usmani I, Timoney N, Laplane C, Gisin N and Afzelius M 2014 {\em New Journal of Physics\/} {\bf 16} 083005 ISSN 1367-2630 (\textit{Preprint} \eprint{1404.3489})

\bibitem{Liu2022}
Liu D~C, Li P~Y, Zhu T~X, Zheng L, Huang J~Y, Zhou Z~Q, Li C~F and Guo G~C 2022 {\em Phys. Rev. Lett.\/} {\bf 129}(21) 210501

\bibitem{Rakonjac2022}
Rakonjac J~V, Corrielli G, Lago-Rivera D, Seri A, Mazzera M, Grandi S, Osellame R and de~Riedmatten H 2022 {\em Science Advances\/} {\bf 8} 2--8 ISSN 2375-2548 (\textit{Preprint} \eprint{2201.03361})

\bibitem{Saglamyurek2015}
Saglamyurek E, Jin J, Verma V~B, Shaw M~D, Marsili F, Nam S~W, Oblak D and Tittel W 2015 {\em Nature Photonics\/} {\bf 9} 83--87 ISSN 1749-4885 (\textit{Preprint} \eprint{1409.0831})

\bibitem{Craiciu2019}
Craiciu I, Lei M, Rochman J, Kindem J~M, Bartholomew J~G, Miyazono E, Zhong T, Sinclair N and Faraon A 2019 {\em Physical Review Applied\/} {\bf 12} 1 ISSN 23317019 (\textit{Preprint} \eprint{1904.08052})

\bibitem{McAuslan2011}
McAuslan D~L, Ledingham P~M, Naylor W~R, Beavan S~E, Hedges M~P, Sellars M~J and Longdell J~J 2011 {\em Phys. Rev. A\/} {\bf 84}(2) 022309

\bibitem{Damon2011}
Damon V, Bonarota M, Louchet-Chauvet A, Chaneli{\`{e}}re T and {Le Gou{\"{e}}t} J~L 2011 {\em New Journal of Physics\/} {\bf 13} 093031 ISSN 1367-2630 (\textit{Preprint} \eprint{1104.4875})

\bibitem{Moiseev2011PRA-PE}
Moiseev S~A 2011 {\em Phys. Rev. A\/} {\bf 83}(1) 012307

\bibitem{Beavan2011}
Beavan S~E, Ledingham P~M, Longdell J~J and Sellars M~J 2011 {\em Opt. Lett.\/} {\bf 36} 1272--1274

\bibitem{Arcangeli2016}
Arcangeli A, Ferrier A and Goldner P 2016 {\em Phys. Rev. A\/} {\bf 93}(6) 062303

\bibitem{Ham2017}
Ham B~S 2017 {\em Scientific Reports\/} {\bf 7} 7655 ISSN 2045-2322

\bibitem{Gerasimov2017OS-ROSE}
Gerasimov K~I, Minnegaliev M~M, Moiseev S~A, Urmancheev R~V, Chaneli{\`{e}}re T and Louchet-Chauvet A 2017 {\em Optics and Spectroscopy\/} {\bf 123} 211--216 ISSN 0030-400X

\bibitem{Liu2024}
Liu J, Liu J, Cui J, Wang L and Zhang G 2024 {\em Optics Express\/} {\bf 32} 6986 ISSN 1094-4087

\bibitem{Liu2020LWG}
Liu C, Zhou Z~Q, Zhu T, Zheng L, Jin M, Liu X, Li P~Y, Huang J~y, Ma Y, Tu T, Yang T~S, Li C~F and Guo G~c 2020 {\em Optica\/} {\bf 7} 192 ISSN 2334-2536 (\textit{Preprint} \eprint{2002.08780})

\bibitem{Minnegaliev2018}
Minnegaliev M~M, Gerasimov K~I, Urmancheev R~V, Moiseev S~A, Chaneli{\`{e}}re T and Louchet-Chauvet A 2018 {Realization of the revival of silenced echo (ROSE) quantum memory scheme in orthogonal geometry} {\em AIP Conference Proceedings\/} vol 1936 p 020012 ISBN 9780735416284 ISSN 15517616

\bibitem{Minnegaliev2023}
Minnegaliev M~M, Gerasimov K~I and Moiseev S~A 2023 {\em JETP Letters\/} {\bf 117} 865--872 ISSN 10906487

\bibitem{Bonarota2014}
Bonarota M, Dajczgewand J, Louchet-Chauvet A, {Le Gou{\"{e}}t} J~L and Chaneli{\`{e}}re T 2014 {\em Laser Physics\/} {\bf 24} 094003 ISSN 1054-660X (\textit{Preprint} \eprint{1311.7331})

\bibitem{Ma2021NLPE}
Ma Y~Z, Jin M, Chen D~L, Zhou Z~Q, Li C~F and Guo G~C 2021 {\em Nature Communications\/} {\bf 12} 1--7 ISSN 20411723 (\textit{Preprint} \eprint{2107.09857})

\bibitem{Hahn1971}
Hahn E, Shiren N and McCall S 1971 {\em Physics Letters A\/} {\bf 37} 265--267 ISSN 03759601

\bibitem{Kaup1977}
Kaup D~J 1977 {\em Phys. Rev. A\/} {\bf 16}(2) 704--719

\bibitem{Silver1985}
Silver M~S, Joseph R~I and Hoult D~I 1985 {\em Phys. Rev. A\/} {\bf 31}(4) 2753--2755

\bibitem{Roos2004}
Roos I and M\o{}lmer K 2004 {\em Phys. Rev. A\/} {\bf 69}(2) 022321

\bibitem{Tian2011}
Tian M, Chang T, Merkel K~D and Randall W 2011 {\em Appl. Opt.\/} {\bf 50} 6548--6554

\bibitem{1998-PRL-BDCZ}
Briegel H~J, D\"ur W, Cirac J~I and Zoller P 1998 {\em Phys. Rev. Lett.\/} {\bf 81}(26) 5932--5935

\bibitem{2007-PRL-Simon}
Simon C, de~Riedmatten H, Afzelius M, Sangouard N, Zbinden H and Gisin N 2007 {\em Physical Review Letters\/} {\bf 98} 190503 ISSN 0031-9007

\bibitem{Sangouard2008}
Sangouard N, Simon C, Zhao B, Chen Y~A, {De Riedmatten} H, Pan J~W and Gisin N 2008 {\em Physical Review A - Atomic, Molecular, and Optical Physics\/} {\bf 77} 1--7 ISSN 10502947 (\textit{Preprint} \eprint{0802.1475})

\bibitem{2012-RMP-Multiphoton}
Pan J~W, Chen Z~B, Lu C~Y, Weinfurter H, Zeilinger A and \ifmmode~\dot{Z}\else \.{Z}\fi{}ukowski M 2012 {\em Rev. Mod. Phys.\/} {\bf 84}(2) 777--838

\bibitem{2001-Nature-DLCZ}
Duan L~M, Lukin M~D, Cirac J~I and Zoller P 2001 {\em Nature\/} {\bf 414} 413--418 ISSN 00280836 (\textit{Preprint} \eprint{0105105})

\bibitem{2021-RevSciInstr-Anwar}
Anwar A, Perumangatt C, Steinlechner F, Jennewein T and Ling A 2021 {\em Review of Scientific Instruments\/} {\bf 92} 041101 ISSN 0034-6748

\bibitem{Bussieres2014}
Bussi{\`{e}}res F, Clausen C, Tiranov A, Korzh B, Verma V~B, Nam S~W, Marsili F, Ferrier A, Goldner P, Herrmann H, Silberhorn C, Sohler W, Afzelius M and Gisin N 2014 {\em Nature Photonics\/} {\bf 8} 775--778 ISSN 17494893 (\textit{Preprint} \eprint{1401.6958})

\bibitem{2023-NatComm-HDR}
Lago-Rivera D, Rakonjac J~V, Grandi S and de~Riedmatten H 2023 {\em Nature Communications\/} {\bf 14} 1889 ISSN 2041-1723

\bibitem{Lago-Rivera2021}
Lago-Rivera D, Grandi S, Rakonjac J~V, Seri A and de~Riedmatten H 2021 {\em Nature\/} {\bf 594} 37--40 ISSN 0028-0836 (\textit{Preprint} \eprint{2101.05097})

\bibitem{2023-OExp-Duranti}
Duranti S, Wengerowsky S, Feldmann L, Seri A, Casabone B and de~Riedmatten H 2024 {\em Optics Express\/} {\bf 32} 26884 ISSN 1094-4087 (\textit{Preprint} \eprint{2307.03509})

\bibitem{2021-NAture-Liu}
Liu X, Hu J, Li Z~F, Li X, Li P~Y, Liang P~J, Zhou Z~Q, Li C~F and Guo G~C 2021 {\em Nature\/} {\bf 594} 41--45 ISSN 0028-0836

\bibitem{Huang2022}
Huang C~X, Hu X~M, Guo Y, Zhang C, Liu B~H, Huang Y~F, Li C~F, Guo G~C, Gisin N, Branciard C and Tavakoli A 2022 {\em Physical Review Letters\/} {\bf 129} 30502 ISSN 10797114

\bibitem{Zeng2024}
Zeng H, Du M~M, Zhong W, Zhou L and Sheng Y~B 2024 {\em Fundamental Research\/} {\bf 4} 851--857 ISSN 26673258

\bibitem{2015-OPtica-Tiranov}
Tiranov A, Lavoie J, Ferrier A, Goldner P, Verma V~B, Nam S~W, Mirin R~P, Lita A~E, Marsili F, Herrmann H, Silberhorn C, Gisin N, Afzelius M and Bussi\`{e}res F 2015 {\em Optica\/} {\bf 2} 279--287

\bibitem{2023-NatComm-Jiang}
Jiang M~H, Xue W, He Q, An Y~Y, Zheng X, Xu W~J, Xie Y~B, Lu Y, Zhu S and Ma X~S 2023 {\em Nature Communications\/} {\bf 14} 1--8 ISSN 20411723 (\textit{Preprint} \eprint{2212.12898})

\bibitem{Saglamyurek2016}
Saglamyurek E, Puigibert M~G, Zhou Q, Giner L, Marsili F, Verma V~B, Nam S~W, Oesterling L, Nippa D, Oblak D and Tittel W 2016 {\em Nature Communications\/} {\bf 7} 1--7 ISSN 20411723 (\textit{Preprint} \eprint{1511.01384})

\bibitem{Zhang2023}
Zhang X, Zhang B, Wei S, Li H, Liao J, Li C, Deng G, Wang Y, Song H, You L, Jing B, Chen F, Guo G and Zhou Q 2023 {\em Science Advances\/} {\bf 9} 1--8 ISSN 23752548 (\textit{Preprint} \eprint{2306.08229})

\bibitem{Sparkes2012}
Sparkes B~M, Hosseini M, Cairns C, Higginbottom D, Campbell G~T, Lam P~K and Buchler B~C 2012 {\em Physical Review X\/} {\bf 2} 021011 ISSN 2160-3308

\bibitem{Wei2022}
Wei S~H, Jing B, Zhang X~Y, Liao J~Y, Yuan C~Z, Fan B~Y, Lyu C, Zhou D~L, Wang Y, Deng G~W, Song H~Z, Oblak D, Guo G~C and Zhou Q 2022 {\em Laser and Photonics Reviews\/} {\bf 16} 1--29 ISSN 18638899 (\textit{Preprint} \eprint{2201.04802})

\bibitem{2022-AVSquantum-Semenenko}
Semenenko V, Hu X, Figueroa E and Perebeinos V 2022 {\em AVS Quantum Science\/} {\bf 4} 012002 ISSN 2639-0213

\bibitem{Moiseev2024_OQM_MC}
Moiseev S~A, Gerasimov K~I, Minnegaliev M~M and Moiseev E~S 2024 Optical quantum memory on macroscopic coherence (\textit{Preprint} \eprint{2408.09991}) \urlprefix\url{https://arxiv.org/abs/2408.09991}

\bibitem{Moiseev2010cavity}
Moiseev S~A, Andrianov S~N and Gubaidullin F~F 2010 {\em Phys. Rev. A\/} {\bf 82}(2) 022311 (\textit{Preprint} \eprint{, Moiseev S., Gubaidullin F., Andrianov S. arXiv: 1001.1140v1, 7 Jan. 2010})

\bibitem{Afzelius2010}
Afzelius M and Simon C 2010 {\em Phys. Rev. A\/} {\bf 82}(2) 022310

\bibitem{Meng2024}
Meng R~R, Liu X, Jin M, Zhou Z~Q, Li C~F and Guo G~C 2024 {\em Chip\/} {\bf 3} 100081 ISSN 27094723

\bibitem{Labonte2024}
Labont\'e L, Alibart O, D'Auria V, Doutre F, Etesse J, Sauder G, Martin A, Picholle E and Tanzilli S 2024 {\em PRX Quantum\/} {\bf 5}(1) 010101

\bibitem{Perminov2018}
Perminov N~S, Tarankova D~Y and Moiseev S~A 2018 {\em Laser Physics Letters\/} {\bf 15} 125203

\bibitem{Moiseev_2017}
Moiseev E~S and Moiseev S~A 2016 {\em Laser Physics Letters\/} {\bf 14} 015202

\bibitem{Matanin2023}
Matanin A~R, Gerasimov K~I, Moiseev E~S, Smirnov N~S, Ivanov A~I, Malevannaya E~I, Polozov V~I, Zikiy E~V, Samoilov A~A, Rodionov I~A and Moiseev S~A 2023 {\em Phys. Rev. Appl.\/} {\bf 19}(3) 034011

\bibitem{Perminov2023}
Perminov N~S and Moiseev S~A 2023 {\em Entropy\/} {\bf 25} 623

\bibitem{Pettit2023}
Pettit R~M, Farshi F~H, Sullivan S~E, Veliz-Osorio A and Singh M~K 2023 {\em Applied Physics Reviews\/} {\bf 10} 031307 ISSN 1931-9401

\end{thebibliography}

\end{document}